\documentclass[aps,prl,reprint,superscriptaddress,floatfix]{revtex4-1}

\usepackage{graphicx}% Include figure files
\usepackage{epstopdf, epsfig}
\usepackage{amsmath}
\usepackage{amssymb}
\usepackage[dvipsnames]{xcolor}
\usepackage{soul}
\usepackage{xr}
\newcommand{\Rey}{\mathit{Re}}

\newcommand{\be}{\begin{equation}}
\newcommand{\ee}{\end{equation}}

\begin{document}

\preprint{}

\title{Resonances in pulsatile channel flow with an elastic wall}% Force line breaks with \\

\author{Duo Xu}
 \email{duo.xu@zarm.uni-bremen.de}
\affiliation{University of Bremen, Center of Applied Space Technology and Microgravity (ZARM), 28359 Bremen, Germany}
\affiliation{Friedrich-Alexander-Universit{\"a}t Erlangen-N{\"u}rnberg, Institute of Fluid Mechanics, 91058 Erlangen, Germany}%
\author{Matthias Heil}
\email{m.heil@maths.manchester.ac.uk}
\affiliation{University of Manchester, School of Mathematics, Manchester, M13 9PL United Kingdom}

\author{Thomas Seeb{\"o}ck}
\affiliation{Friedrich-Alexander-Universit{\"a}t Erlangen-N{\"u}rnberg, Institute of Fluid Mechanics, 91058 Erlangen, Germany}%

\author{Marc Avila}
\email{marc.avila@zarm.uni-bremen.de}
\affiliation{University of Bremen, Center of Applied Space Technology and Microgravity (ZARM), 28359 Bremen, Germany}
\affiliation{Friedrich-Alexander-Universit{\"a}t Erlangen-N{\"u}rnberg, Institute of Fluid Mechanics, 91058 Erlangen, Germany}%
\affiliation{University of Bremen, MAPEX Center for Materials and Processes, 28359 Bremen, Germany}

\date{\today}% It is always \today, today,

\begin{abstract}
Interactions between fluids and elastic solids are ubiquitous
in application ranging from aeronautical and civil engineering to
physiological flows. Here we study the pulsatile flow through a
{two-dimensional} Starling resistor as a simple model for unsteady flow in elastic
vessels. We {numerically} solve the equations governing the flow and the
large-displacement elasticity and show that the system responds as a
forced harmonic oscillator with non-conventional damping. We derive an
analytical prediction for the amplitude of the oscillatory wall
deformation, and thus the conditions under which resonances occur or
vanish.
 \end{abstract}

\maketitle

\begin{figure}
\includegraphics[width=1\linewidth]{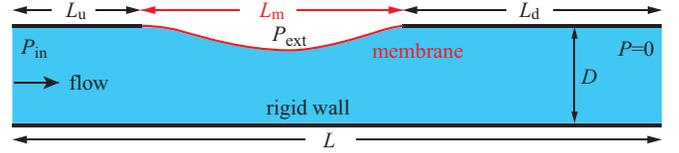}
	\caption{\label{fig:setup}(Color online) Sketch of fluid flow
          in a collapsible channel. A pulsatile pressure $P_\text{in}$
          of mean $P_0$ and frequency $\Omega$ drives fluid of
          kinematic viscosity $\nu$ and density $\rho$ through a
          channel of total length $L$ and width $D$.
          The lower channel wall is rigid,
          whereas a pre-stressed, elastic membrane of length
          $L_\text{m}$ is clamped between two rigid segments at the
          upper wall and is pressurized by an external pressure
          $P_\text{ext}$.  The upstream and downstream segments have
          lengths $L_\text{u}$ and $L_\text{d}$, respectively.}
\end{figure}

% plays an important role in many
%applications. 
Flow-induced pressure fluctuations acting
on elastic structures can excite large-amplitude oscillations 
-- the most famous example being the
catastrophic failure of the Tacoma-Narrows bridge \cite{billah1991}.
In physiology, fluid-structure interaction is associated with 
cardiovascular disease~\cite{Ku97}, but it also helps regulate the blood supply
to internal organs~\cite{Shapiro77} and return blood to the
heart during diastole~\cite{Casey08}. Physiological flows are extremely complex
and feature a large variability across individuals, which prevents accurate
predictions even with state-of-the-art computational methods~\cite{Valen-Sendstad2018}.
Therefore there has been a desire to study the key physical mechanisms in simpler
setups~\cite{Heil11}. The  Starling
resistor~\cite{Knowlton12,Conrad69} is a
canonical system that has been widely used to investigate the
nonlinearly coupled dynamics of fluid flow and the deformation of
elastic vessels. The setup consists of an elastic tube mounted between
two rigid pipes in a pressurized chamber (or its two-dimensional
analogue, the collapsible channel shown in Fig.~\ref{fig:setup}). 

In Starling resistors the flow is typically driven by a constant
pressure drop between inlet and outlet. For this case, rich
nonlinear phenomena, such as flow limitation \cite{Kamm79} and
self-excited oscillations
\cite{Conrad69,Jensen03,Bertram08,Stewart09}, have been observed. By
contrast, studies of pulsatile flows through elastic tubes and
channels {\cite{Conrad69,Low91,Tubaldi2016,tsigklifis2017,Stelios19,amabili2020b}} are comparatively
scarce despite the pulsatile nature of blood flows. A notable exception is
Amabili \emph{et al.}'s recent study~\cite{Amabili20} in which part of
an excised human aorta was mounted between two rigid pipes
and subjected to physiological pulsatile pressure and flow rates.

In this \emph{Letter} we show that pulsatile flow in a {two-dimensional} collapsible channel exhibits strong resonances, 
reminiscent of a forced damped harmonic
oscillator. Guided by this observation, we develop a simple mathematical 
model which successfully predicts the resonances, the
phase lag between the amplitude and the imposed pressure, and also the
conditions under which resonances vanish.

{We consider a fluid of kinematic viscosity $\nu$ and density $\rho$, whose motion is governed by the incompressible
Navier--Stokes equations}
\begin{equation}
\dfrac{\partial {\bf u}}{\partial t} + {\bf u} \cdot \nabla {\bf
u} = -\nabla  p + \nabla^2  {\bf u},\quad
\nabla \cdot {\bf u}= 0.
\label{eq:NSeqn}
\end{equation}
Here and elsewhere all lengths are scaled on the channel height, $D$,
and time, {$t$,} on the timescale for viscous diffusion, $D^2/\nu$. The fluid
velocity {${\bf u}$} is non-dimensionalized on $\nu/D$ and the pressure, {$p$,} on
$\rho\nu^2/D^2$. We set the pressure at the downstream end of the
channel to zero and drive the flow by setting the dimensionless
pressure at the upstream end to
\begin{equation}
\label{inflow_bc}
p = p_{\rm in}(t)=12 \Rey \,l \left(1 + A\sin\,(\alpha^2 \, t)\right),
\end{equation}
where $l=L/D$. The dimensionless forcing frequency $\alpha^2 = \Omega D^2/\nu$ {($\Omega$ is the dimensional
  frequency and $\alpha$ the Womersley number)}
characterizes the ratio of the
timescale for viscous diffusion to the period of the imposed pressure
pulsation; 
$A$ is the amplitude of the oscillatory component
of the pressure relative to the steady one. The Reynolds number $\Rey
= D{\cal U}/\nu$ is defined with the mean speed of the Poiseuille flow
generated by a steady pressure drop $P_0$ in the
undeformed channel, ${\cal U} = P_0 D^2/(12 \rho \nu L)$. The boundary
conditions for the velocity are no-slip on the walls, and parallel
flow is assumed at the inflow and outflow boundaries. 

We model the elastic segment {of the wall} as a thin, massless membrane (of
dimensional thickness $\mathfrak{h}$ and Young's modulus $E$, subject
to a dimensional pre-stress $\Sigma_0$) which deforms in response to
the combined effects of the external pressure and the fluid
{stresses.} The resulting traction {vector} acting on the membrane,
non-dimensionalized on {the pre-stress $\Sigma_0$,} is given by
{\begin{equation}
\label{eq:nondimloadvector}
{\bf f}=-p_\text{ext} {\bf n} + \frac{1}{T} \left[ p {\bf n} -
  \left(\nabla {\bf u} + (\nabla {\bf u})^{\rm T}\right) \cdot {\bf n} \right],
\end{equation}}
where ${\bf n}$ is the outer normal to the membrane,
{$p_\text{ext}=P_\text{ext}/\Sigma_0$ and the superscript $^{\rm T}$ denotes 
the transpose of a matrix.} The
parameter $T= \Sigma_0\,D^2/(\rho\nu^2)$ represents the ratio of the
pre-stress to the fluid pressure and is a measure of the tension in the bounding
membrane. We parametrize the shape of the membrane by a dimensionless
Lagrangian coordinate $\xi$ so that the position vector
to a material point in the membrane is
given by ${\bf R}(\xi,t) = {\bf r}(\xi) + {\bf d}(\xi,t)$. Here {${\bf
	r}(\xi) = [\xi, 1]^{\rm T}$} defines the undeformed configuration and $
{\bf d}(\xi,t)$ is the displacement vector. The membrane
deformation is governed by the principle of virtual displacements
{\begin{equation}
\label{eq:virtualdisplacement}
\int_{0}^{l_\text{m}} \left((\sigma_0 + \gamma)\delta \gamma
+ \frac{1}{12}h^2\kappa \,\delta\kappa -
\frac{\sigma_0\Lambda}{h} \ {\bf f} \cdot \delta {\bf R} \right) d\xi = 0,
\end{equation}}
where $h=\mathfrak{h}/D$ is the dimensionless thickness of the membrane,
{$\sigma_0=\Sigma_0/E$ the dimensionless pre-stress,}
$\gamma=\partial d_x/\partial {\xi} + \frac{1}{2}[(\partial d_x/\partial
{\xi})^2 + (\partial d_y/\partial {\xi})^2]$ is {a 
measure of the extensional strain,} and
$\kappa = [(\partial^2 d_y/\partial {\xi^2})(1+\partial d_x/\partial
{\xi}) - (\partial^2 d_y/\partial {{\xi}^2}) (\partial d_y/\partial
{\xi})]/\Lambda$ {provides a measure of the bending deformation,}
with $\Lambda=[(1+\partial
d_x/\partial \xi)^2+(\partial d_y/\partial {\xi})^2]^{1/2}$. 
{Both measures  are fully geometrically nonlinear. The only
linearization occurs in the assumption of incrementally linear Hookean
behavior in the constitutive equation, which is based on the assumption that
the pre-stress is much larger than the stresses induced by the actual
deformation,  $\sigma_0 \gg 1$.} % Comment 5, Referee 1

We solved the time-dependent fully-coupled fluid-structure interaction
problem with the open-source library {\tt oomph-lib}
\cite{heil2006}.  All simulations shown here were performed with
$A=1$, $h=0.01$, $l_\text{m}=L_\text{m}/D=10$, $l_\text{u}=L_{\rm
  u}/D=5$ and $l_\text{d}=L_\text{d}/D=10$. 

\begin{figure}[!htb]
\centering
\includegraphics[width=1.0\linewidth]{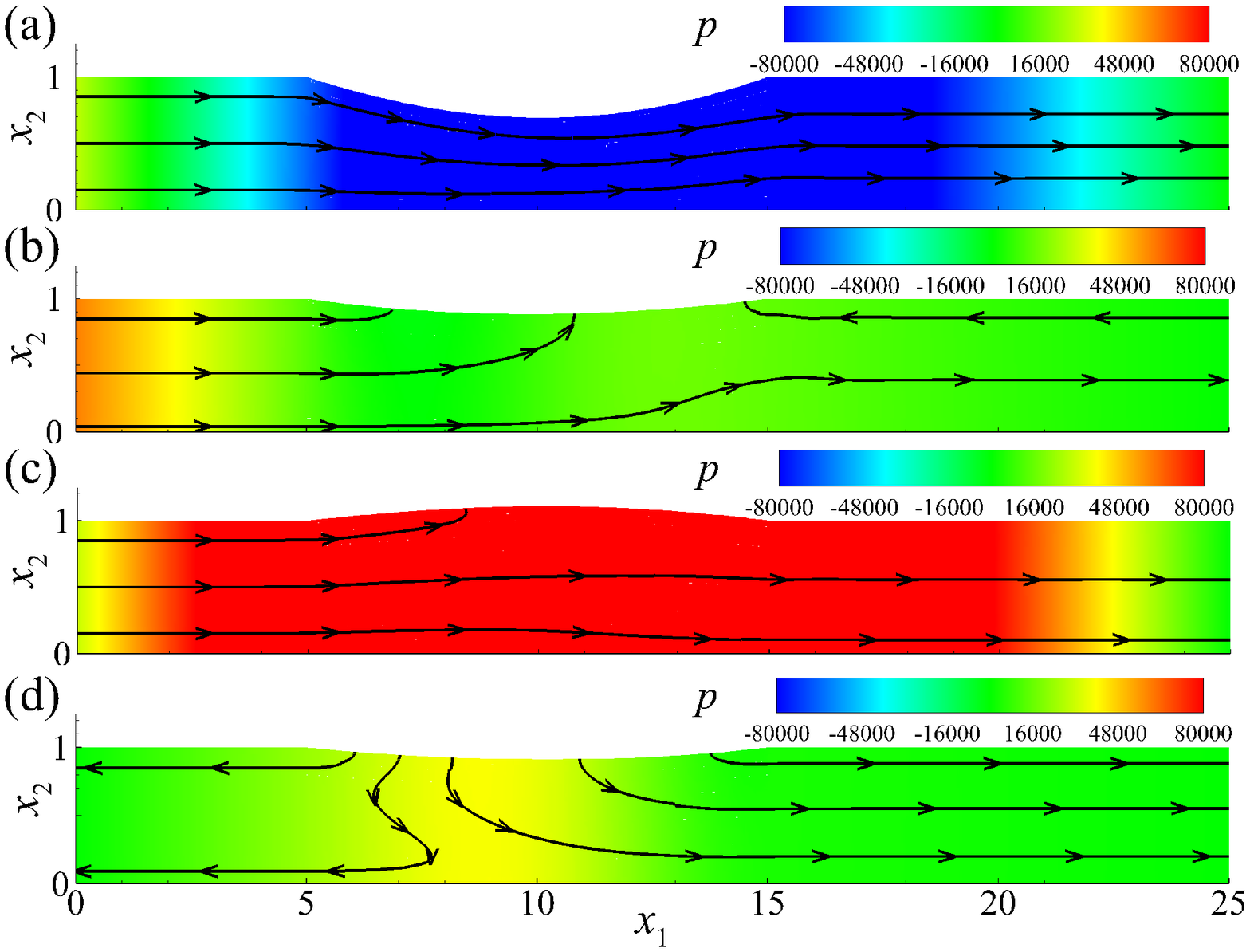}\\
\includegraphics[width=0.49\linewidth, trim={0.3cm 0cm 1.5cm 0.5cm},clip]{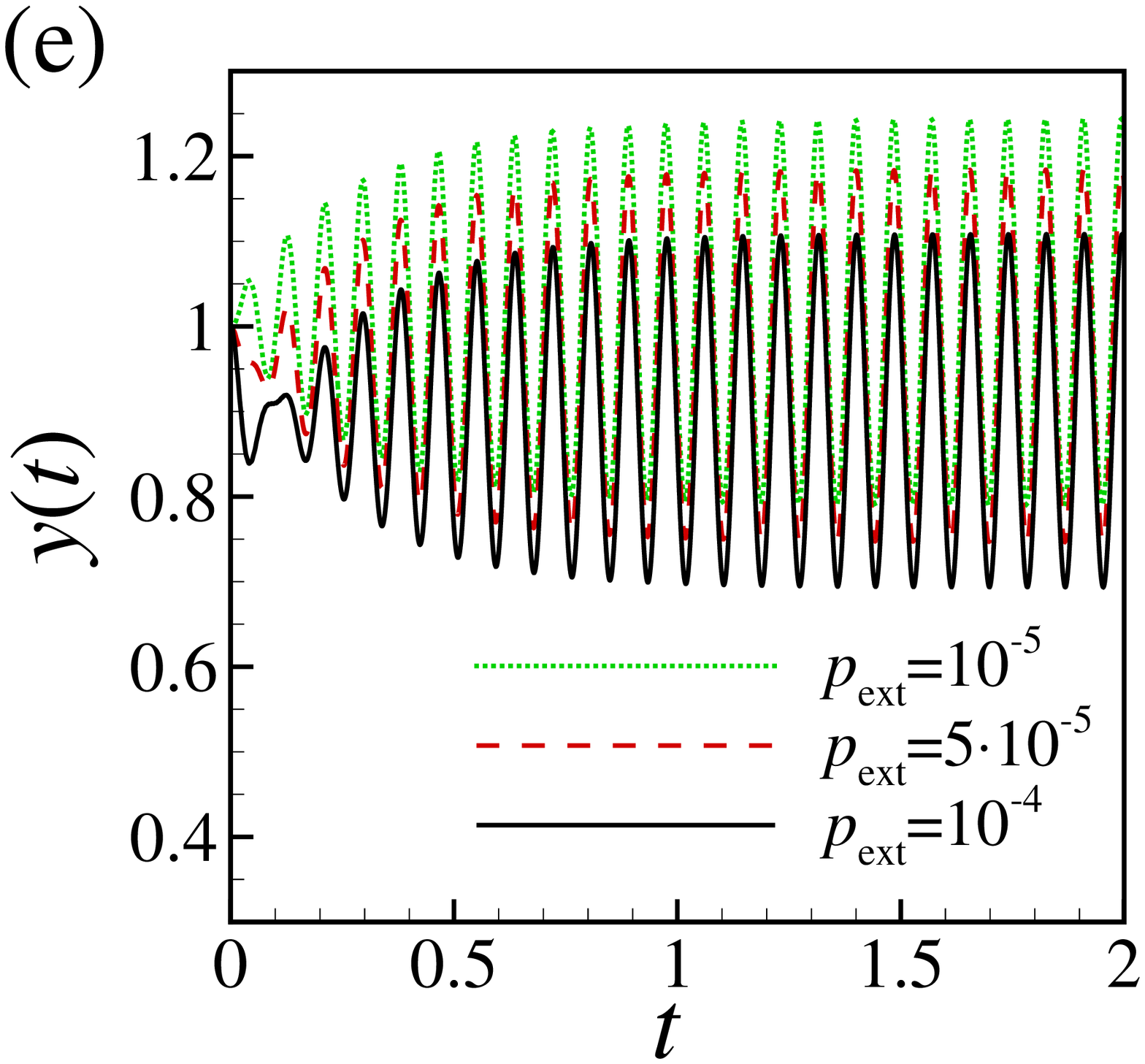}
\includegraphics[width=0.49\linewidth, trim={0.3cm 0cm 1.5cm 0.5cm}, clip]{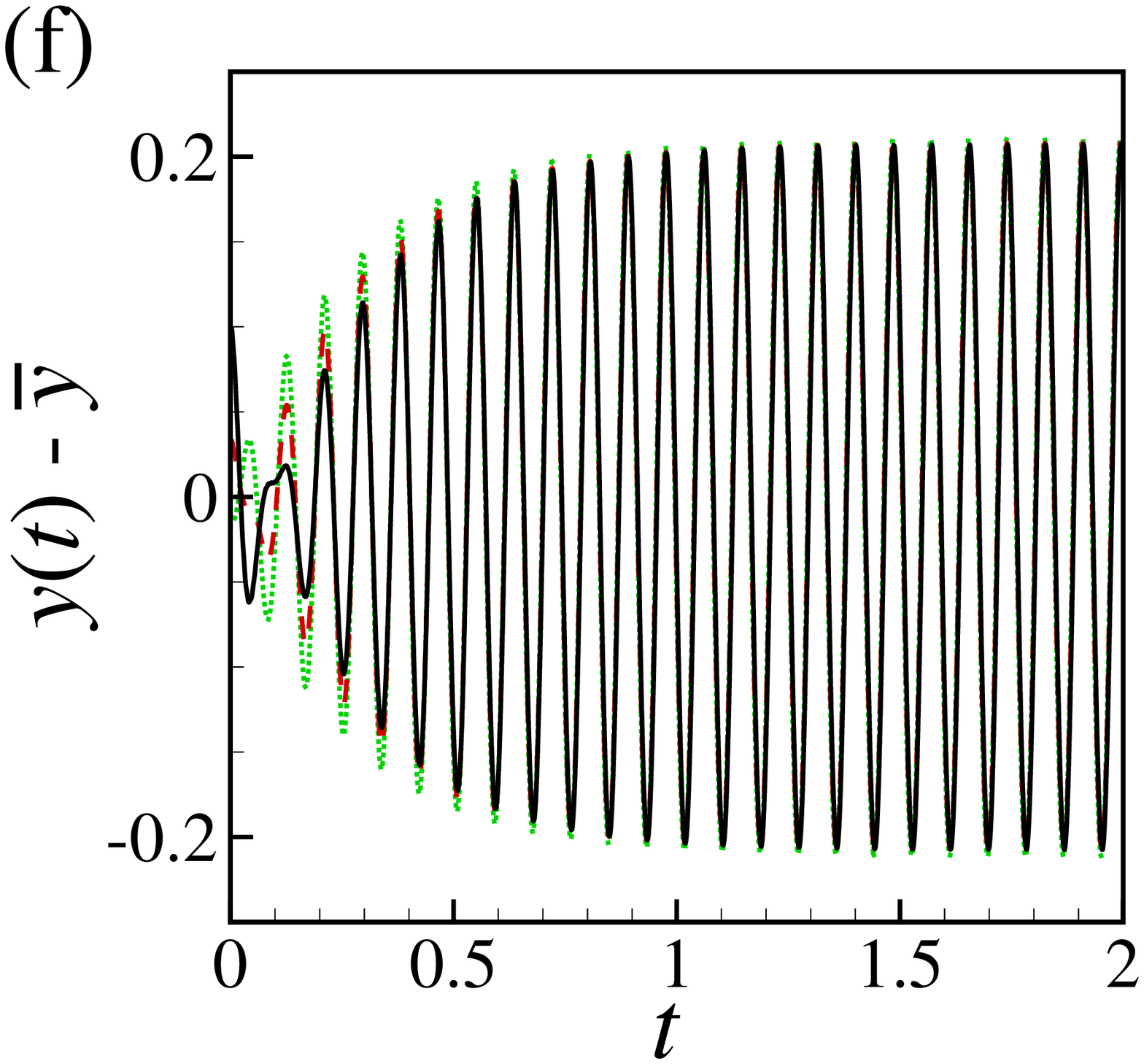}
\includegraphics[width=1\linewidth, trim={1cm 0cm 2cm 0cm}, clip]{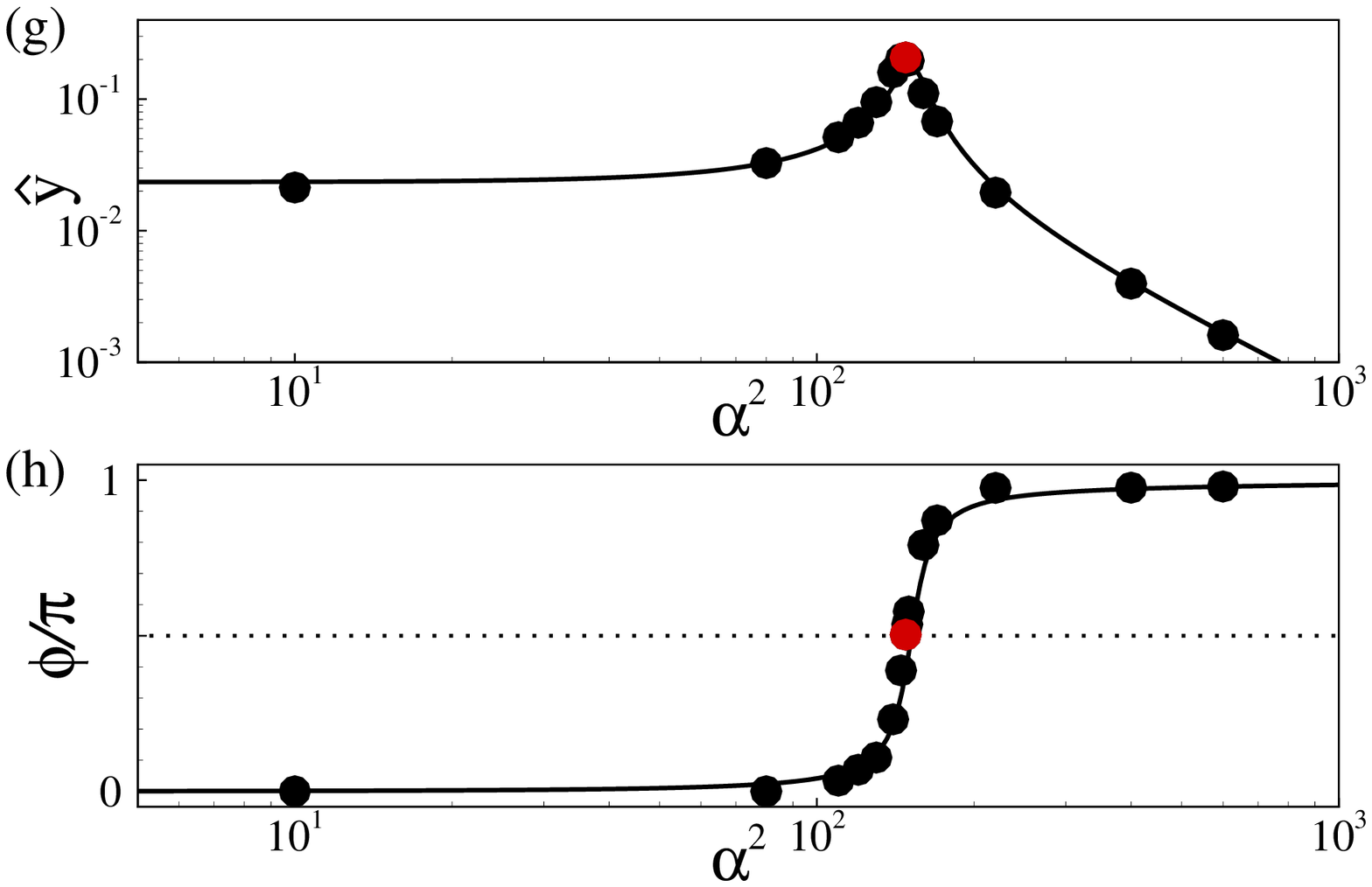}
\caption{\label{fig:flow_fields_and_displacement} (Color online)
  (a)-(d) Pressure contours and streamlines at four equally-spaced
  instants throughout the period of the oscillation for
  $\alpha^2=148$, $\Rey=100$, $\sigma_0=10^3$, $T=10^9$ and
  {$p_\text{ext}=10^{-4}$} after the decay of the initial transients. (e)
  Time trace of the vertical displacement of the membrane mid-point,
  $y(t)$, at the same parameters as (a-d) except for three different
  external pressures {($p_\text{ext}=10^{-5}$, $5\cdot 10^{-5}$ and $10^{-4}$).} (f) The
  same data of (e) after subtracting the time-averaged displacements,
  $\overline{y}$, following the decay of the initial transients
  ($\overline{y} = 1.017, 0.965, 0.900$, respectively). (g) The
  oscillation amplitude and (h) the corresponding phase lag between
  response and forcing as a function of $\alpha^2$, where the red dots
  mark the case for (a)--(d) {and the horizontal dotted line marks $\phi=\pi/2$.} 
  All other parameters are as in  (a)--(d).}
\end{figure}

We started the simulations from an initial condition in which the
membrane is undeformed and the velocity field is steady Poiseuille
flow. Following the decay of initial transients  the 
system settles into a time-periodic motion
with the period of the forcing,
$2\pi/\alpha^2$. The snapshots in Figs.~\ref{fig:flow_fields_and_displacement}(a-d)
show that the inward wall motion displaces a significant amount of fluid from the central
region of the channel and thus creates strong \emph{sloshing} flows
which are superimposed on the pressure-driven {pulsatile} flow. These are
reminiscent of the flows observed in a study of self-excited oscillations 
in collapsible channels~\cite{Jensen03}.

We characterize the dynamics of the system by monitoring the vertical
displacement of the membrane at its midpoint, $y(t)$.
Fig.~\ref{fig:flow_fields_and_displacement}(e) shows time traces of
this quantity for a range of external pressures. In
Fig.~\ref{fig:flow_fields_and_displacement}(f) we plot the same data
but subtract the time-average displacement $\overline{y}$ following
the decay of the initial transients (whose duration is of the
order of the viscous time unit, $D^2/\nu$). We observe that the amplitude of
the steady-state oscillations, $\widehat{y}$, is approximately
independent of the external pressure (and from now on we set
{$p_\text{ext}=10^{-4}$).}  Fig.~\ref{fig:flow_fields_and_displacement}(g) shows that the amplitude
of the oscillations, $\widehat{y}$, exhibits a sharp maximum at a
specific forcing frequency, $\alpha^2_{\rm max}$. Furthermore, the
phase lag $\phi$ between the displacement $y(t)$ and the forcing
pressure $p_{\rm in}(t)$ displays a $90^{\circ}$ phase shift when the
amplitude reaches its maximum; see Fig.~\ref{fig:flow_fields_and_displacement}(h). 

\begin{figure}
\centering
  \includegraphics[width=1\linewidth, trim={0cm 0cm 0cm 0cm},
    clip]{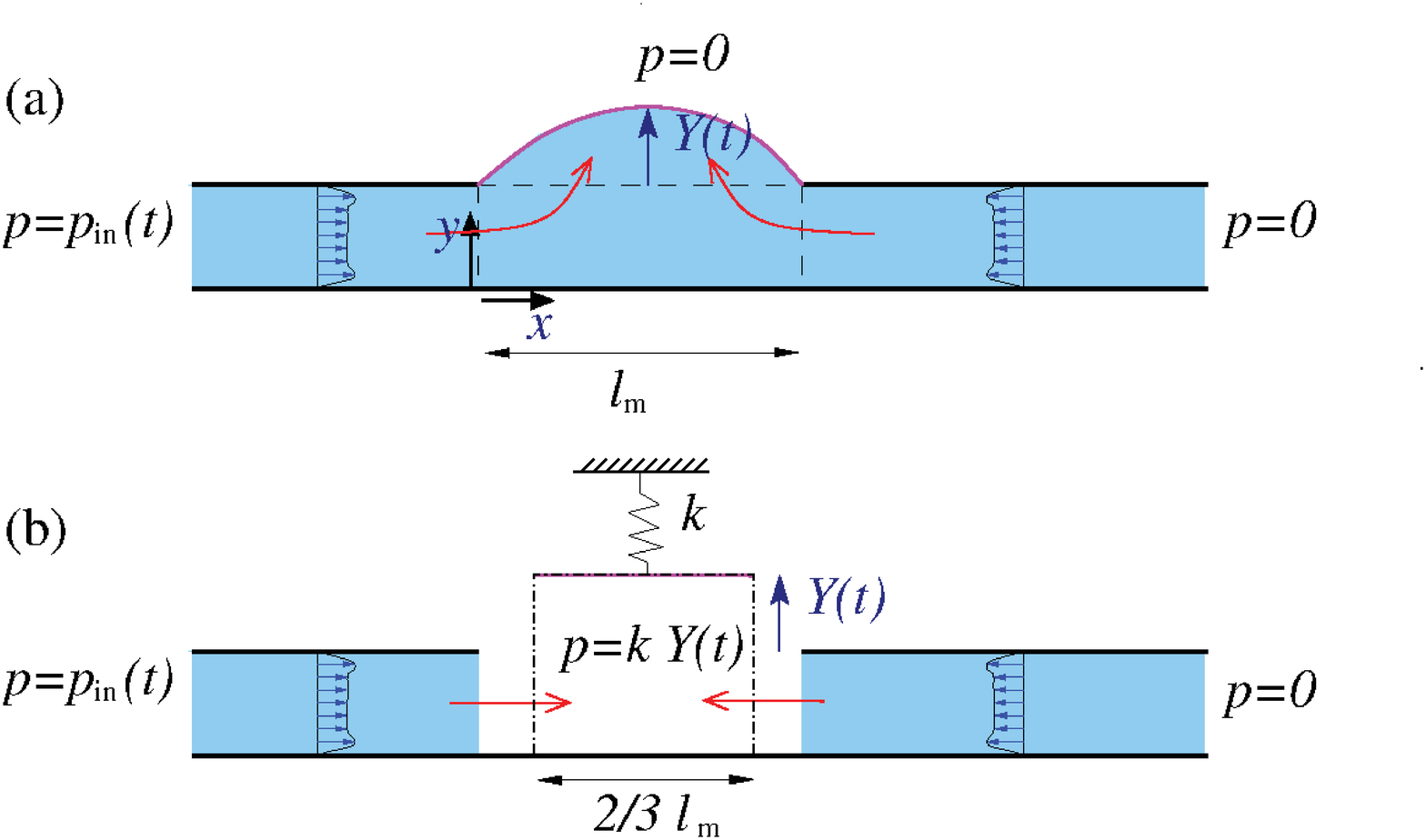}
    \includegraphics[width=1\linewidth, trim={0.75cm 0cm 1cm 0cm},
      clip]{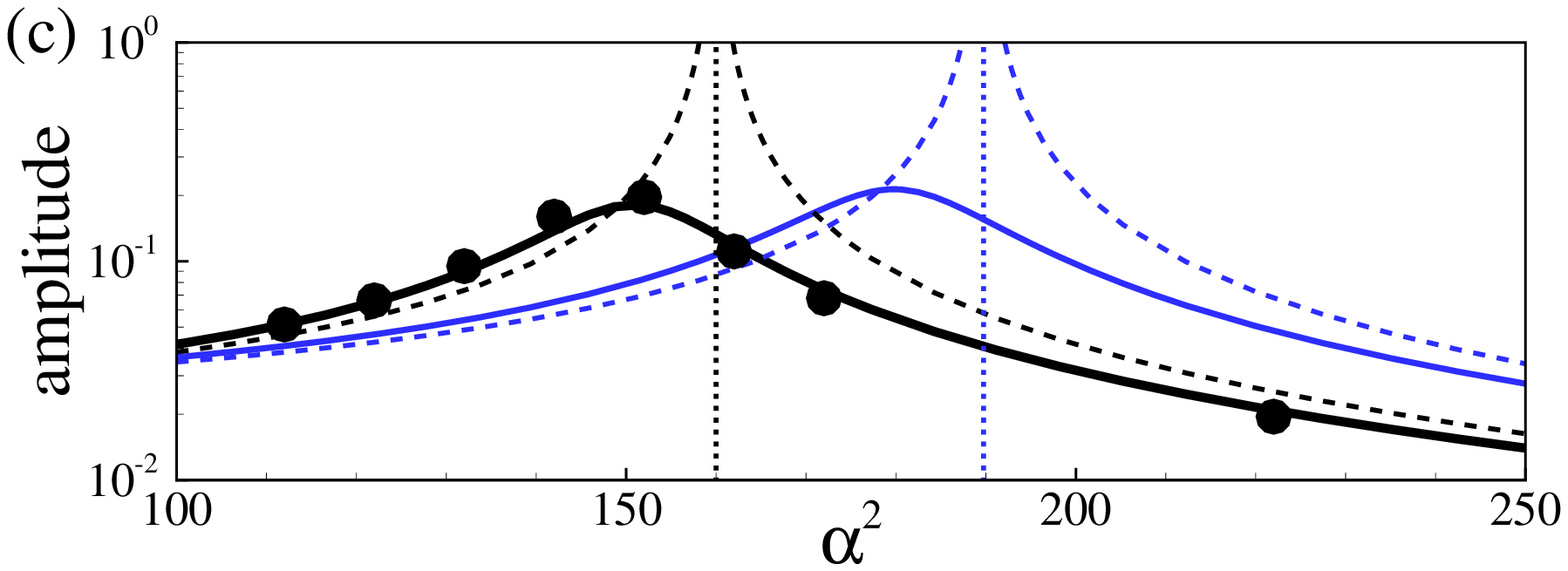}
\caption{\label{fig:model_sketch}(Color online) (a)--(b) Sketch of the
  model. (c) Oscillation amplitude against $\alpha^2$ from   the model
  at $Re=100$, $\sigma_0=10^3$, $T=10^9$, where the solid   and the
  dashed lines denote the viscous and the inviscid prediction,   and
  black and blue correspond to $\beta=0.25$ and $\beta=0$. The   black
  symbols show the results from the simulations   (as
  in Fig.~\ref{fig:flow_fields_and_displacement}(g), with some
  symbols omitted for clarity). The  dotted lines mark the corresponding
  eigenfrequencies.}
\end{figure}

To elucidate the mechanism responsible for this behavior we will now
develop a simple theoretical model that describes the response of the
collapsible channel to the imposed pressure pulsations at its
upstream end. Since we found the external pressure to have little
effect on the system's behavior, we set it to zero and thus consider
the setup sketched in Fig. \ref{fig:model_sketch}(a).  We assume the
upstream and downstream rigid parts of the channel to be sufficiently
long, $l_{\rm u}, l_{\rm d} \gg 1$, so that in these segments
{the horizontal component of
the velocity, $u$, is much larger than its vertical counterpart.} Our
computations show that this assumption is appropriate even in the
relatively short channels used in our simulations; see
Fig.~\ref{fig:flow_fields_and_displacement}(a)--(d). The horizontal
component of the momentum equation (\ref{eq:NSeqn}) can then be
approximated by 
\be
\label{long_wavelength_momentum}
\frac{\partial u}{\partial t} = -\frac{\partial p}{\partial x} +
\frac{\partial^2 u}{\partial y^2},
\ee
where the pressure gradient only depends on time,
$\partial p/\partial x = G(t)$, and we have $u = u(y,t)$.
We assume that the vertical displacement of the elastic membrane
can be described by the product of a mode shape $M(x)$ and an
amplitude $Y(t)$, so that
\be
y_{\rm m}(x,t) = 1+ Y(t)\, M(x).
\ee
Based on the shapes observed in the computations, we approximate
$M(x)$ by a quadratic function, $M(x) = 4 (x/l_{\rm m})(1-(x/l_{\rm
  m}))$. Given that the elastic membrane is under a large,
approximately constant tension
we describe its deformation by Laplace's law,
implying that the fluid pressure in the elastic segment is given by
the product of the membrane curvature and its tension. For the assumed
mode shape the dimensionless fluid pressure under the membrane is $p_{\rm m} = k \, Y(t),$
where $k = 8 \,h \,T/l_{\rm m}^2$. By exploiting that the flows in the two rigid segments are fully developed and coupled by mass conservation, we show in the Supplementary Material that the displacement of the membrane $Y$ obeys the following equation
\be
\label{eigen_equation}
\frac{2}{3}l_{\rm m} \frac{{\rm d}^2 Y}{{\rm d}t^2} +
k \frac{l_{\rm u} + l_{\rm d}}{l_{\rm u} l_{\rm d}} \ Y
+ \left.
\left( \frac{\partial u_{\rm d}}{\partial y} -
       \frac{\partial u_{\rm u}}{\partial y} \right)
\right|_{y=0}^1 = \frac{p_{\rm in}(t)}{l_{\rm u}}.
\ee
{This equation can be interpreted in terms of the 
difference in the pressure gradients in the upstream and downstream
segments
driving an acceleration in the net flow away from the centre, which
must be balanced by the change in volume of the elastic section, see 
eq.~(S13) in the Supplementary Material.}

The last term on the left hand side of eq.~\eqref{eigen_equation} arises from the
viscous terms in the momentum equation and represents the effect of
the viscous shear stresses acting on the walls of the rigid
segments. The remaining terms show that in the absence of viscous
damping the system is a forced linear oscillator with {natural frequency}
\be
\label{theoretical_alpha_eig}
\alpha_{\rm eig}^2 = \left( 
\frac{3}{2} \frac{k}{l_{\rm m}}
\frac{l_{\rm u} + l_{\rm d}}{l_{\rm u} l_{\rm d}}
\right)^{1/2}
=\left( 
\frac{12\,h\, T (l_{\rm u} + l_{\rm d})} 
{l_{\rm m}^3l_{\rm u} l_{\rm d}}
\right)^{1/2}.
\ee
The damping term in equation (\ref{eigen_equation}) is more complicated
than in a standard harmonic oscillator (see the Supplementary Material). 

Our model equation \eqref{eigen_equation} therefore predicts that the collapsible channel
behaves like the linear oscillator sketched in
Fig.~\ref{fig:model_sketch}(b): The elastic membrane of length $l_{\rm
  m}$ is equivalent to a piston of width $\frac{2}{3}l_{\rm m}$,
mounted on a spring of stiffness $k$. The piston is displaced by the
net influx of fluid from the rigid segments and sets the fluid
pressure acting at their internal boundaries. The system's
oscillations are governed by a dynamic balance between fluid inertia
and the elastic restoring forces, with the fluid viscosity providing
damping. 

The amplitude $|\widehat{Y}|$ of the time-harmonic solutions, $Y(t) = \widehat{Y} \exp({\rm
  i}\alpha^2 t)$, to (\ref{eigen_equation}) is given in the Supplementary Material, eq.~(S11). 
The blue solid line in Fig.~\ref{fig:model_sketch}(c) shows a plot of
the theoretically predicted amplitude as a function of
the forcing frequency $\alpha^2$ for the same parameters as in
Fig.~\ref{fig:flow_fields_and_displacement}(g). The thin blue dashed
line shows the corresponding inviscid response, with the
{natural frequency} $\alpha^2_{\rm eig}$ shown by the blue vertical dotted line. 
Viscous effects eliminate the unbounded
response of the inviscid system at $\alpha^2= \alpha^2_{\rm eig}$ and reduce the resonant frequency. The
theoretical predictions are in good qualitative agreement with the
computational results, but they over-estimate
the resonant frequency. This is a consequence of us having neglected
the dynamics of the fluid that moves within the elastic segment
itself. We can include this effect by replacing $l_{\rm
  u/d}$ by corresponding effective lengths $l_{\rm u/d}^{\rm [eff]} =
l_{\rm u/d} + \beta l_{\rm m}$, where the parameter $\beta$ represents
the fraction of the fluid in the elastic segment that participates in
the oscillatory (sloshing) motion.  The black lines in
Fig. \ref{fig:model_sketch}(c) show the theoretical predictions for $\beta = 1/4$. 
This value produces near-perfect agreement
with the results of the simulations for all the cases considered 
{(see Table~S1 in the Supplementary Material for a full list of all computations)} 
and is kept fixed hereinafter.

\begin{figure}
\includegraphics[width=0.49\linewidth, trim={0.5cm 0cm 0.25cm 0cm}, clip]{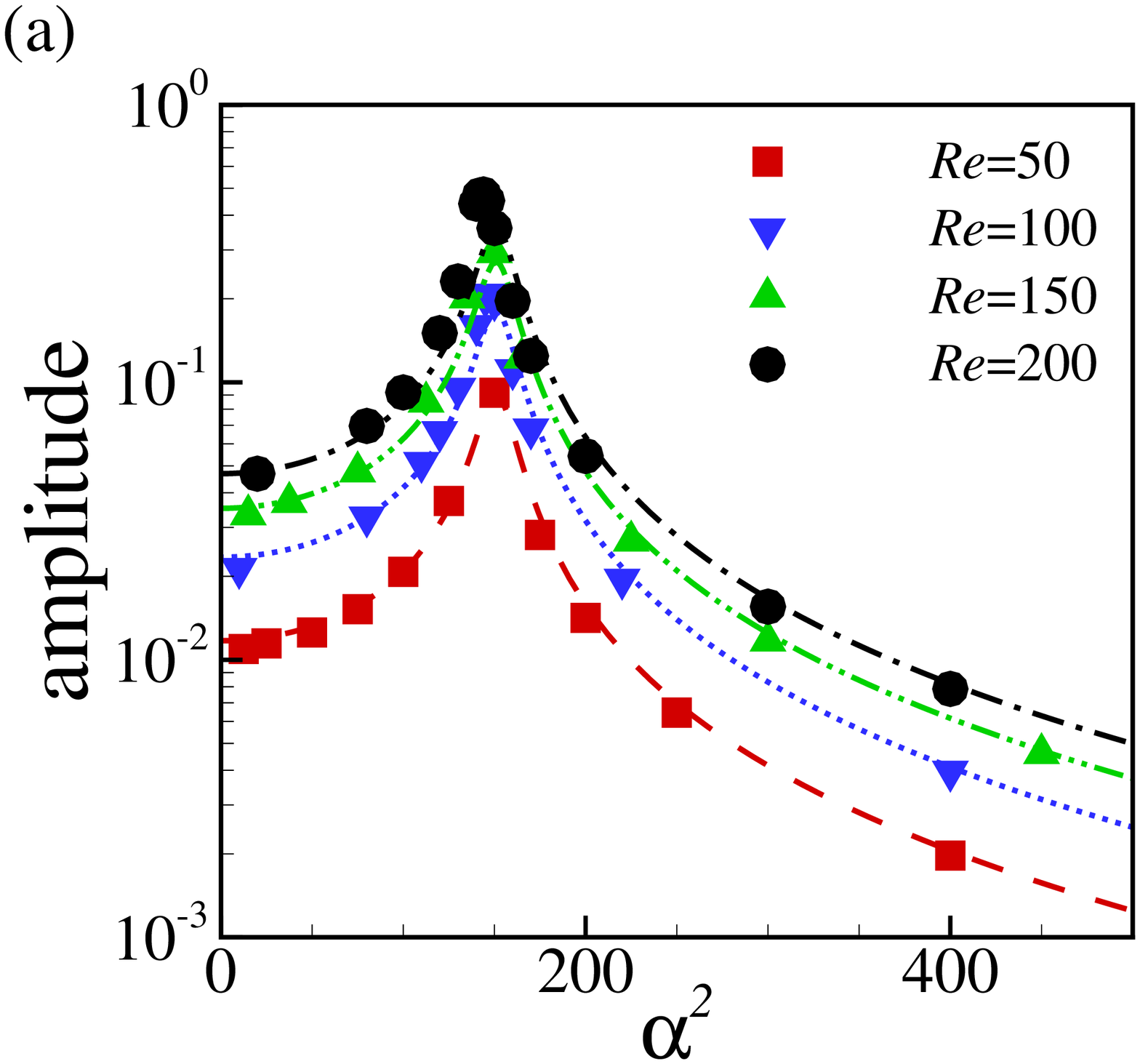}    
\includegraphics[width=0.49\linewidth, trim={0.5cm 0cm 0.25cm 0cm}, clip]{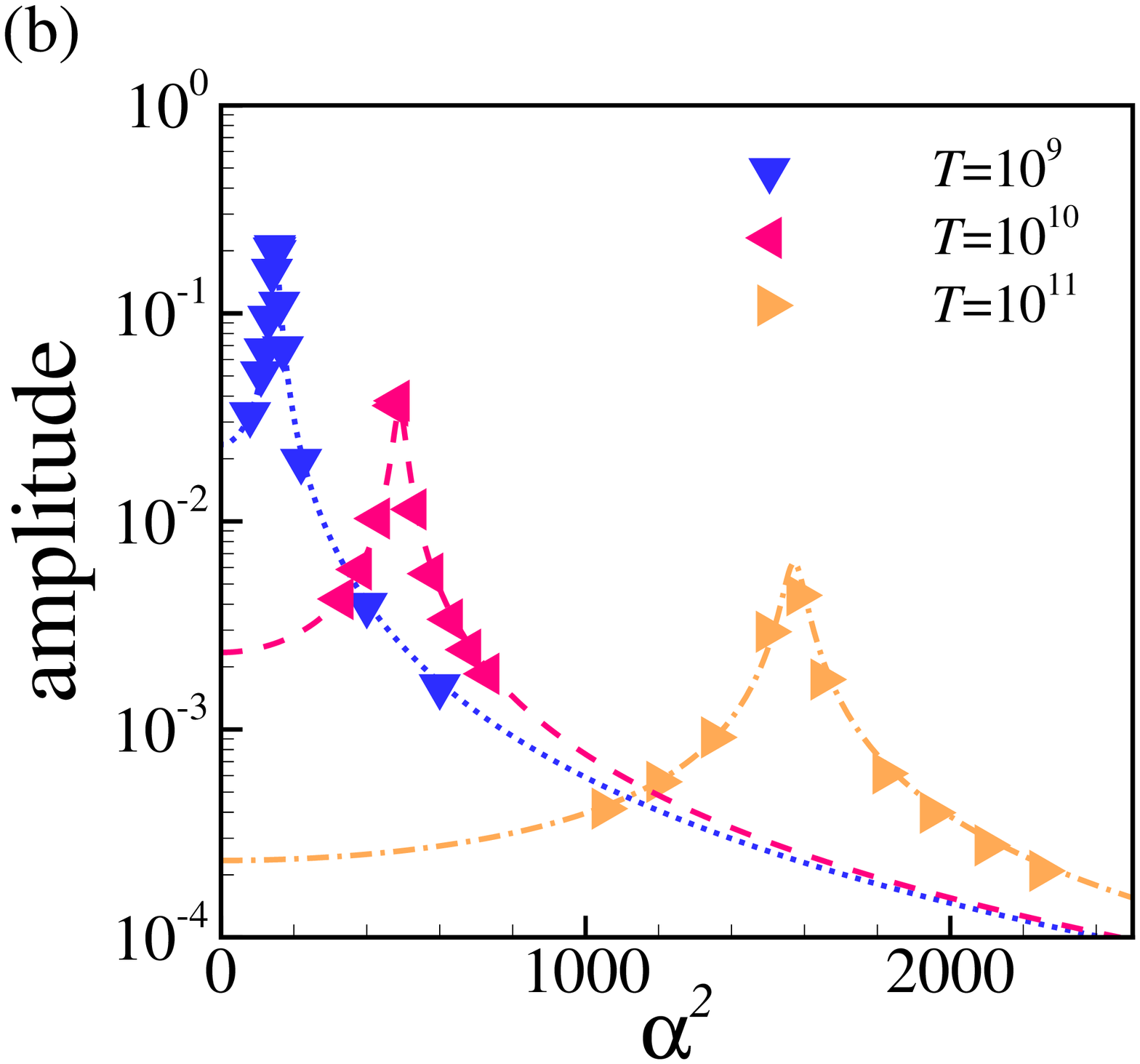}  
\includegraphics[width=0.49\linewidth, trim={0.5cm 0cm 0.25cm 0cm}, clip]{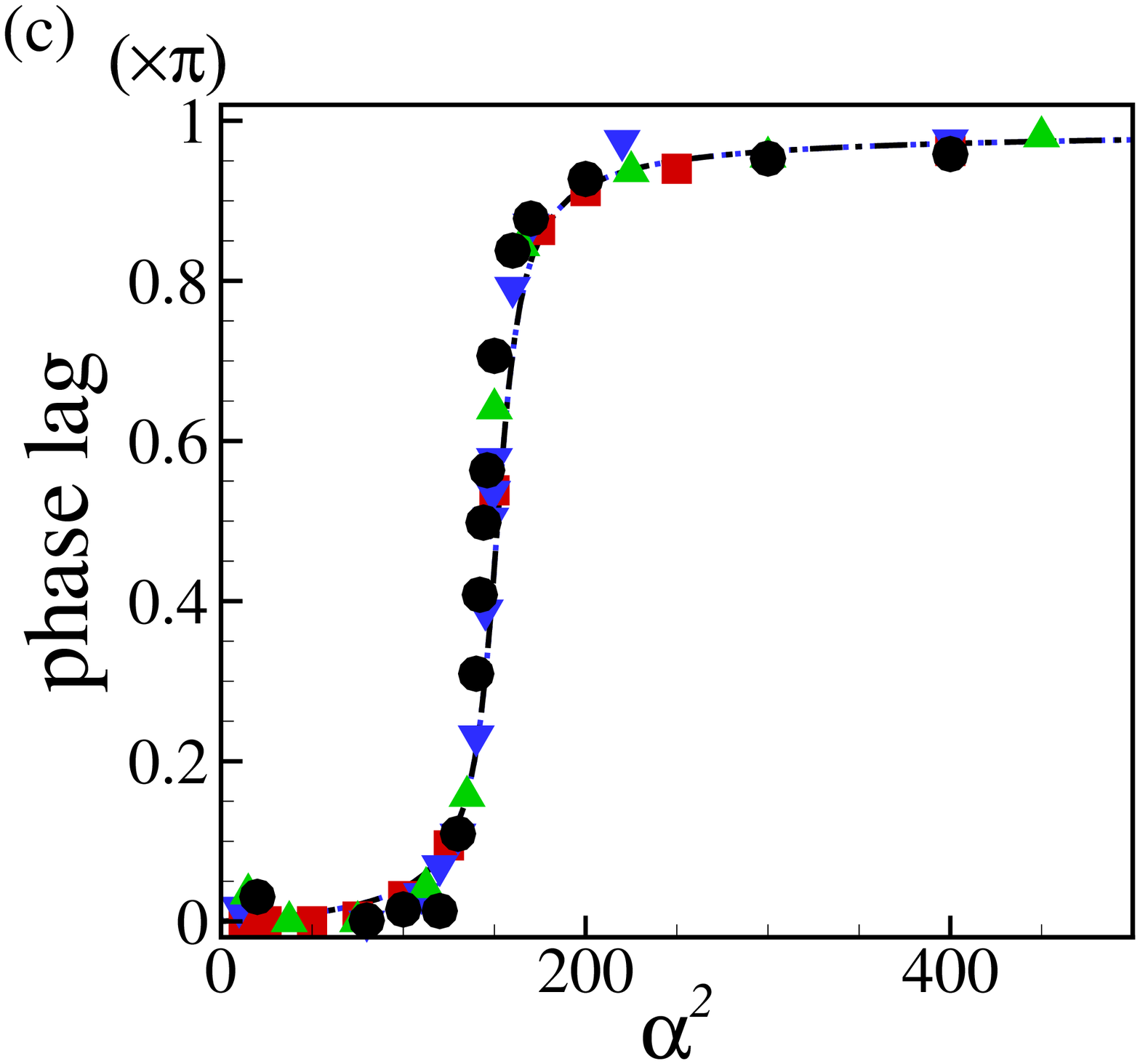}  
\includegraphics[width=0.49\linewidth, trim={0.5cm 0cm 0.25cm 0cm}, clip]{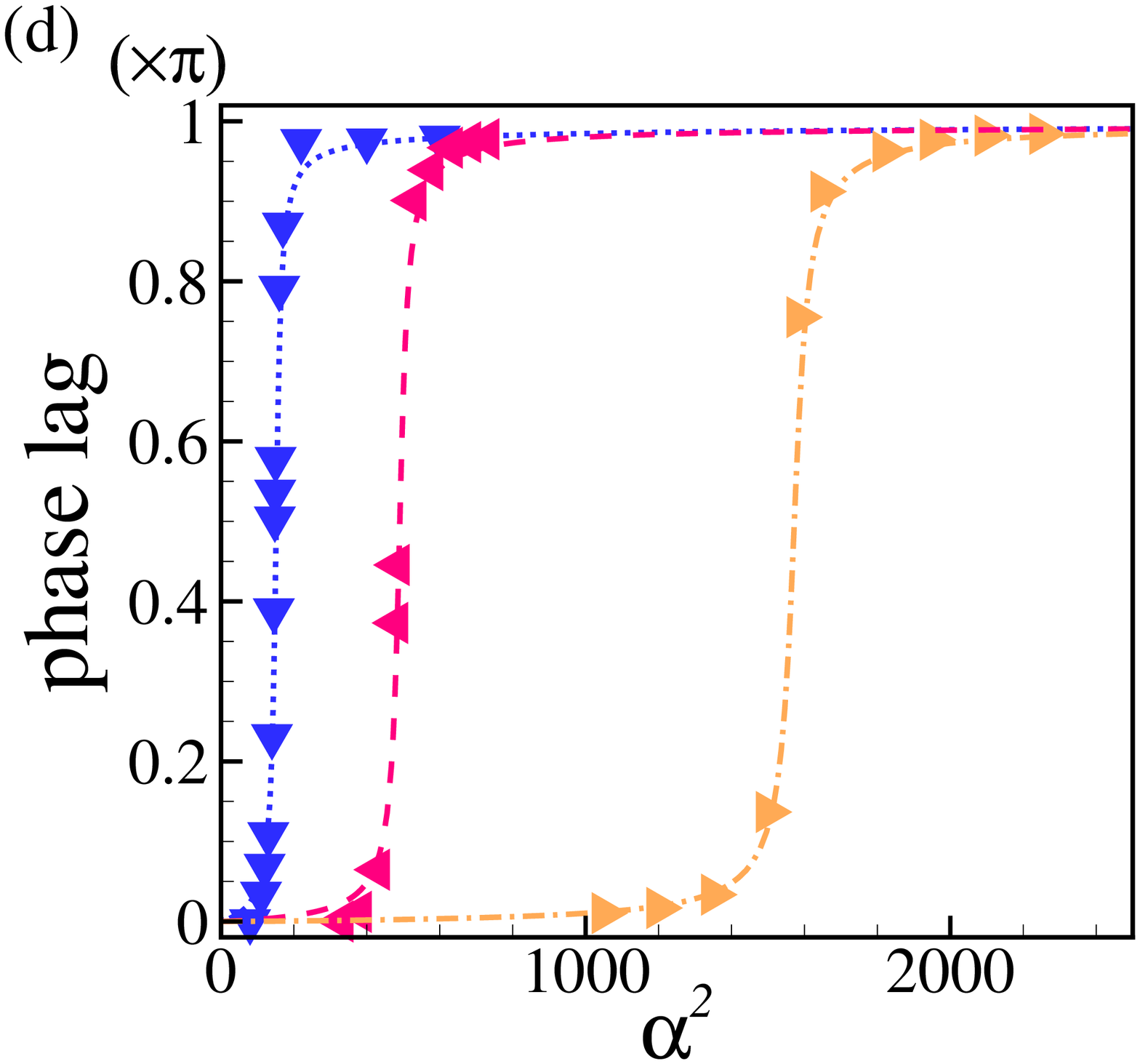}
\caption{\label{fig:amplitude_phase} (Color online) (a) Oscillation amplitude against $\alpha^2$ for $\sigma_0=10^3$ and $T=10^9$ for four Reynolds numbers, and the corresponding phase lag (c). (b) Oscillation amplitude against $\alpha^2$ for $Re=100$, $\sigma_0=10^3$ and three values of $T$, and the corresponding phase lag (d). The symbols and the lines denote the simulation and the model predictions, respectively.}
\end{figure}

We note that the theoretical model predicts the system's response
to be controlled by the parameter $T$ and the geometry; the
Reynolds number $Re$ is
predicted to affect only the amplitude of the response, but not the
{resonant} frequency. 
Fig.~\ref{fig:amplitude_phase} shows the amplitude
(top row) and phase (bottom row) as a function of the forcing frequency $\alpha^2$ for a constant value of $T$ (left) and a constant
value of $Re$ (right). The agreement between the theoretical
predictions and computational results  is remarkable,
even for oscillations of large amplitude (Fig.~\ref{fig:amplitude_phase} 
includes cases where the amplitude reaches values as large as 48\% of the channel width). 
The amplitude increases proportionally to  the
Reynolds number. A reduction in the membrane tension ($T$)
increases the amplitude of the oscillation and reduces the resonant
frequency. This suggests that for sufficiently small values of $T$ the
maximum amplitude may occur in the quasi-steady limit ($\alpha^2 \to
0$). However, for the parameter values of Fig.
\ref{fig:amplitude_phase} this happens when the theoretically
predicted amplitude exceeds the undeformed channel width, i.e.
$|\widehat{Y}| > 1$, rendering the theoretical model inapplicable.
To explore the disappearance of the resonance at smaller
values of $T$ we therefore reduced the Reynolds number
significantly.

\begin{figure}
\centering \includegraphics[width=0.49\linewidth, trim={0.5cm 0cm 0.25cm 0cm}, clip]{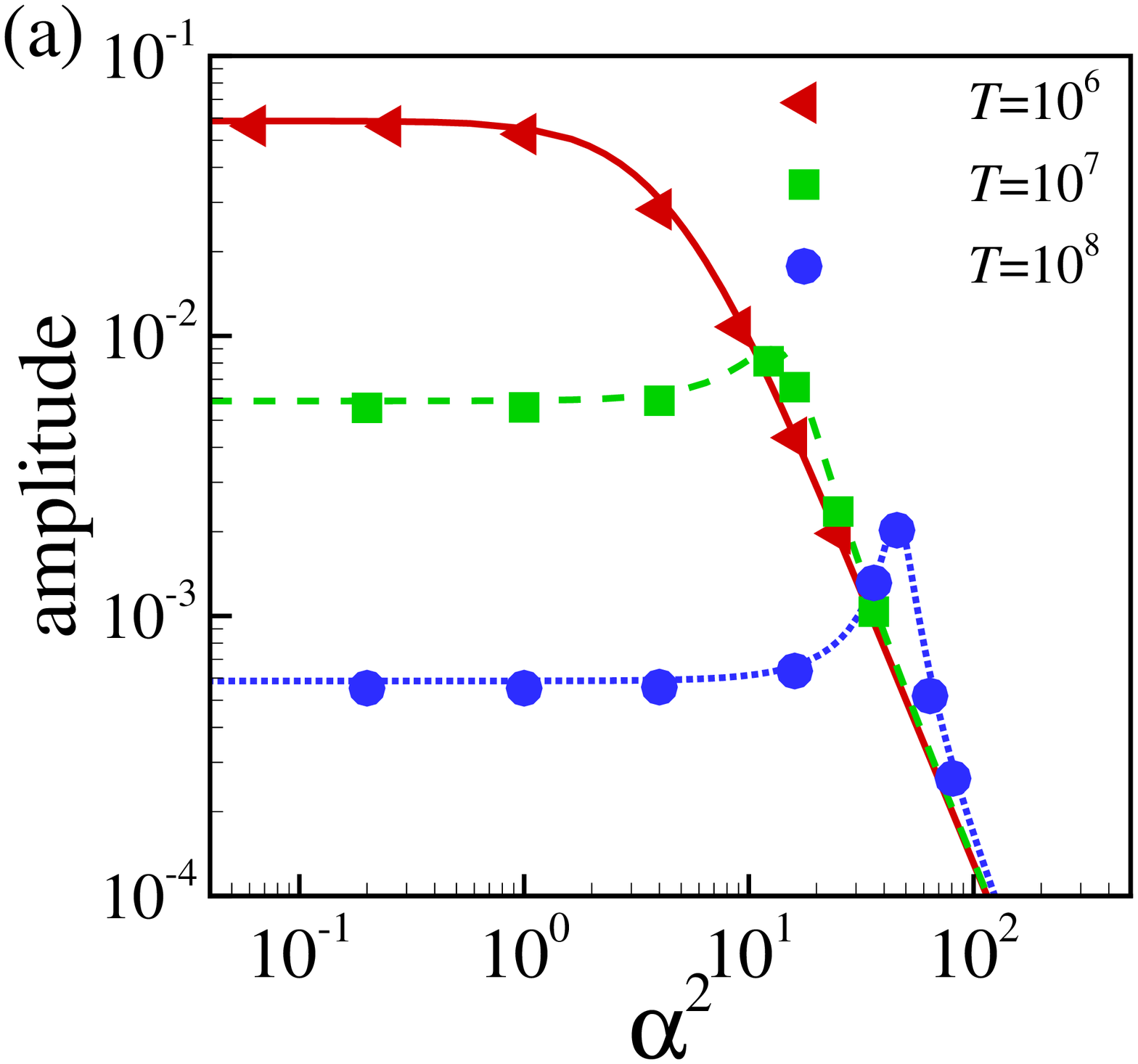} \centering
\includegraphics[width=0.49\linewidth, trim={0.5cm 0cm 0.25cm 0cm},
  clip]{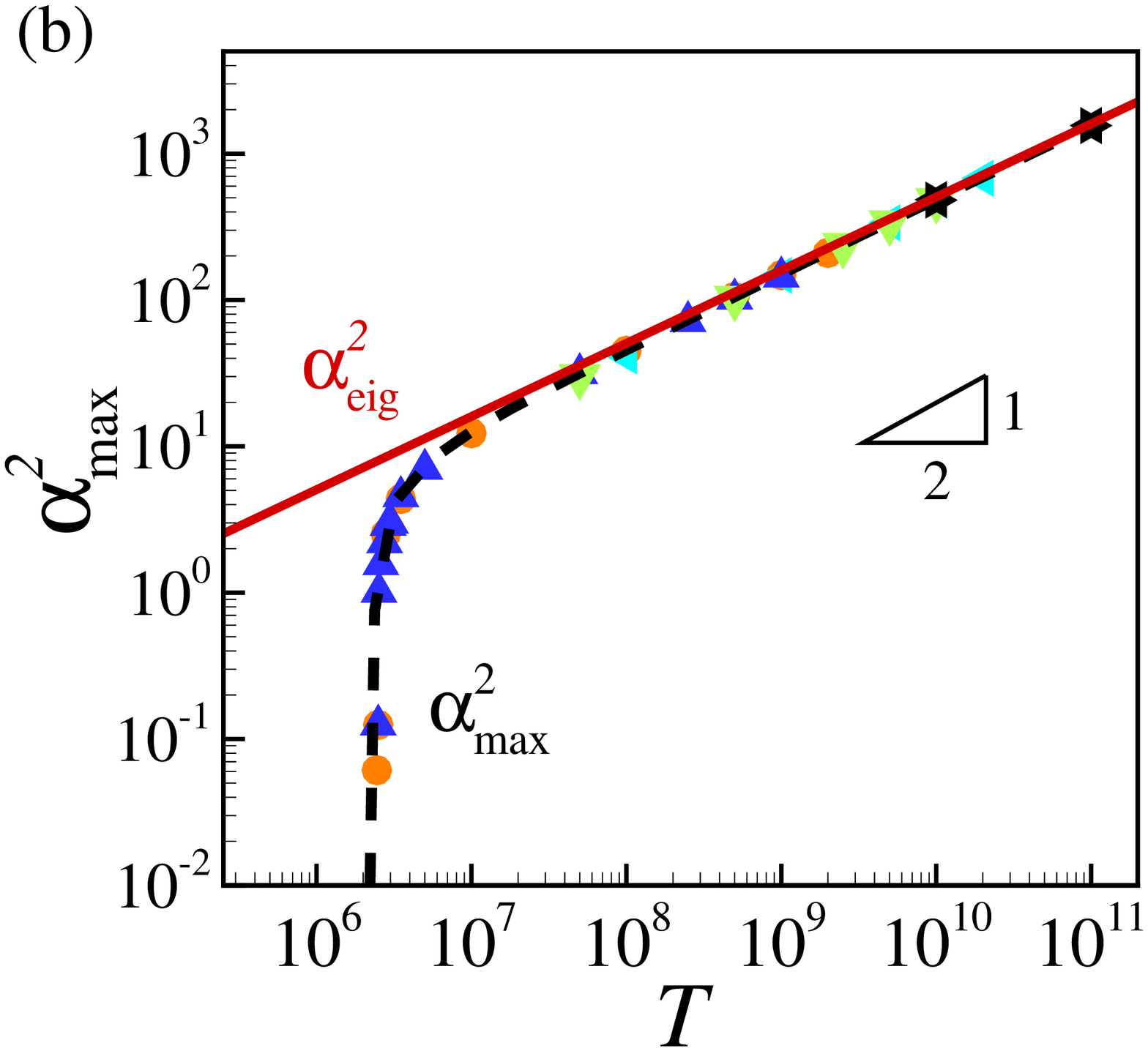}
\caption{\label{fig:eigenfrequency}(Color online) (a) Oscillation
  amplitude against frequency $\alpha^2$ at $Re=0.25$,
  $\sigma_0=10^3$. The maximum oscillation amplitude for $T=10^8$ and
  $10^7$ occurs at $\alpha^2\approx45$ and $12.5$, respectively, and
  at $T=10^6$ the maximum amplitude is approached as $\alpha^2\to
  0$. The symbols and the lines denote the simulation and the model,
  respectively. (b) The frequency $\alpha^2_{\text{max}}$ as a
  function of $T$. The symbols denote $\alpha^2_{\text{max}}$ from the
  simulation. The red solid line and the black dashed line denote the
  model predictions of $\alpha^2_{\text{eig}}$ and
  $\alpha^2_{\text{max}}$, respectively. The parameters used in the
  simulations are listed in table~S1 in the Supplementary Material.}
\end{figure}

Fig.~\ref{fig:eigenfrequency}(a) shows a plot of the amplitude as a
function of the forcing frequency $\alpha^2$ for three values of
$T$ and for a Reynolds number of $Re=0.25$. For $T=10^8$ there is a
clearly defined resonance at $\alpha_{\rm max}^2
\approx 45$; a reduction of $T$ to $10^7$ increases the
maximum amplitude but weakens the resonance and moves it to smaller
values of the forcing frequency, $\alpha_{\rm max}^2 \approx 12.5$;
finally, for $T=10^6$ the maximum amplitude is obtained in
the quasi-steady limit, implying the disappearance of the
resonance. 

Fig.~\ref{fig:eigenfrequency}(b) shows how the {natural frequency,}
$\alpha^2_{\rm eig}$, and the frequency, $\alpha^2_{\rm max}$, at
which the system has its maximum response, depend on the system
parameters. For large values of $T$ the maximum response occurs close
to the {natural frequency}, $\alpha^2_{\rm max} \lesssim \alpha^2_{\rm
  eig}$, and both scale with the square root of $T$ as suggested by
equation~(\ref{theoretical_alpha_eig}). Just below $T \approx 2\times10^6$ the
resonance disappears. Finally, we probed the dependence of the system's response on
the geometry by performing simulations for various 
combinations of $l_{\rm m}$, $l_{\rm u}$, $l_{\rm d}$ and $h$. As
shown in Fig.~S1 in the Supplementary Material, the scaling suggested by the
inviscid approximation \eqref{theoretical_alpha_eig} leads to a
near-perfect collapse of all the results onto a single master curve
{($\beta=0.25$ was kept fixed).} % Main Comment Referee 1

In summary,  the response of a collapsible channel 
is described by a harmonic
oscillator with non-standard damping, even in regimes where the
imposed pulsations in fluid pressure induce very large wall deflections.
Our model accurately predicts the response of the system
as a function of the ten independent parameters that govern it
($l_{\rm m}$, $l_{\rm u}$, $l_{\rm d}$, $h$, $p_\text{ext}$, $Re$,
$A$, $\alpha^2$, $T$ and $\sigma_0$). The tension and the dimensions of the
channel segments solely determine the system's {natural frequency,}
whereas the amplitude of the response also depends on the frequency and amplitude  of
the pressure pulsations (the latter is set by $A\,Re$, see
Fig.~S2 in the Supplementary Material). While our simulations 
were performed for a 2D system the
mechanism can be generalized to a 3D setting.
The characterization of oscillations during which the elastic {tube} %segment
undergoes an axisymmetric inflation is straightforward, whereas the characterization of non-axisymmetric oscillations
could benefit from a `tube-law'-based description~\cite{whittaker2010}.

\begin{acknowledgments}
This work was supported by the Deutsche Forschungsgemeinschaft (DFG) in the framework of the research unit FOR 2688 `Instabilities, Bifurcations and Migration in Pulsatile Flows' under grant AV 120/6-1. D.X. gratefully acknowledges the support from Alexander von Humboldt Foundation (3.5-CHN/1154663STP).
\end{acknowledgments}
\nocite{Walters2017,Mandre2010}
%\bibliography{FSI}

\begin{thebibliography}{22}%
\makeatletter
\providecommand \@ifxundefined [1]{%
 \@ifx{#1\undefined}
}%
\providecommand \@ifnum [1]{%
 \ifnum #1\expandafter \@firstoftwo
 \else \expandafter \@secondoftwo
 \fi
}%
\providecommand \@ifx [1]{%
 \ifx #1\expandafter \@firstoftwo
 \else \expandafter \@secondoftwo
 \fi
}%
\providecommand \natexlab [1]{#1}%
\providecommand \enquote  [1]{``#1''}%
\providecommand \bibnamefont  [1]{#1}%
\providecommand \bibfnamefont [1]{#1}%
\providecommand \citenamefont [1]{#1}%
\providecommand \href@noop [0]{\@secondoftwo}%
\providecommand \href [0]{\begingroup \@sanitize@url \@href}%
\providecommand \@href[1]{\@@startlink{#1}\@@href}%
\providecommand \@@href[1]{\endgroup#1\@@endlink}%
\providecommand \@sanitize@url [0]{\catcode `\\12\catcode `\$12\catcode
  `\&12\catcode `\#12\catcode `\^12\catcode `\_12\catcode `\%12\relax}%
\providecommand \@@startlink[1]{}%
\providecommand \@@endlink[0]{}%
\providecommand \url  [0]{\begingroup\@sanitize@url \@url }%
\providecommand \@url [1]{\endgroup\@href {#1}{\urlprefix }}%
\providecommand \urlprefix  [0]{URL }%
\providecommand \Eprint [0]{\href }%
\providecommand \doibase [0]{http://dx.doi.org/}%
\providecommand \selectlanguage [0]{\@gobble}%
\providecommand \bibinfo  [0]{\@secondoftwo}%
\providecommand \bibfield  [0]{\@secondoftwo}%
\providecommand \translation [1]{[#1]}%
\providecommand \BibitemOpen [0]{}%
\providecommand \bibitemStop [0]{}%
\providecommand \bibitemNoStop [0]{.\EOS\space}%
\providecommand \EOS [0]{\spacefactor3000\relax}%
\providecommand \BibitemShut  [1]{\csname bibitem#1\endcsname}%
\let\auto@bib@innerbib\@empty
%</preamble>
\bibitem [{\citenamefont {Billah}\ and\ \citenamefont
  {Scanlan}(1991)}]{billah1991}%
  \BibitemOpen
  \bibfield  {author} {\bibinfo {author} {\bibfnamefont {K.~Y.}\ \bibnamefont
  {Billah}}\ and\ \bibinfo {author} {\bibfnamefont {R.~H.}\ \bibnamefont
  {Scanlan}},\ }\href@noop {} {\bibfield  {journal} {\bibinfo  {journal}
  {American Journal of Physics}\ }\textbf {\bibinfo {volume} {59}},\ \bibinfo
  {pages} {118} (\bibinfo {year} {1991})}\BibitemShut {NoStop}%
\bibitem [{\citenamefont {Ku}(1997)}]{Ku97}%
  \BibitemOpen
  \bibfield  {author} {\bibinfo {author} {\bibfnamefont {D.~N.}\ \bibnamefont
  {Ku}},\ }\href@noop {} {\bibfield  {journal} {\bibinfo  {journal} {Annual
  Review of Fluid Mechanics}\ }\textbf {\bibinfo {volume} {29}},\ \bibinfo
  {pages} {399} (\bibinfo {year} {1997})}\BibitemShut {NoStop}%
\bibitem [{\citenamefont {Shapiro}(1977)}]{Shapiro77}%
  \BibitemOpen
  \bibfield  {author} {\bibinfo {author} {\bibfnamefont {A.~H.}\ \bibnamefont
  {Shapiro}},\ }\href@noop {} {\bibfield  {journal} {\bibinfo  {journal}
  {Journal of Biomechanical Engineering}\ }\textbf {\bibinfo {volume} {99}},\
  \bibinfo {pages} {126} (\bibinfo {year} {1977})}\BibitemShut {NoStop}%
\bibitem [{\citenamefont {Casey}\ and\ \citenamefont {Hart}(2008)}]{Casey08}%
  \BibitemOpen
  \bibfield  {author} {\bibinfo {author} {\bibfnamefont {D.~P.}\ \bibnamefont
  {Casey}}\ and\ \bibinfo {author} {\bibfnamefont {E.~C.}\ \bibnamefont
  {Hart}},\ }\href@noop {} {\bibfield  {journal} {\bibinfo  {journal} {The
  Journal of Physiology}\ }\textbf {\bibinfo {volume} {586}},\ \bibinfo {pages}
  {5045} (\bibinfo {year} {2008})}\BibitemShut {NoStop}%
\bibitem [{\citenamefont
  {Valen-Sendstad~\emph{et~al.}}(2018)}]{Valen-Sendstad2018}%
  \BibitemOpen
  \bibfield  {author} {\bibinfo {author} {\bibfnamefont {K.}~\bibnamefont
  {Valen-Sendstad~\emph{et~al.}}},\ }\href {\doibase
  10.1007/s13239-018-00374-2} {\bibfield  {journal} {\bibinfo  {journal}
  {Cardiovascular Engineering and Technology}\ }\textbf {\bibinfo {volume}
  {9}},\ \bibinfo {pages} {544} (\bibinfo {year} {2018})}\BibitemShut {NoStop}%
\bibitem [{\citenamefont {Heil}\ and\ \citenamefont {Hazel}(2011)}]{Heil11}%
  \BibitemOpen
  \bibfield  {author} {\bibinfo {author} {\bibfnamefont {M.}~\bibnamefont
  {Heil}}\ and\ \bibinfo {author} {\bibfnamefont {A.~L.}\ \bibnamefont
  {Hazel}},\ }\href@noop {} {\bibfield  {journal} {\bibinfo  {journal} {Annual
  Review of Fluid Mechanics}\ }\textbf {\bibinfo {volume} {43}},\ \bibinfo
  {pages} {141} (\bibinfo {year} {2011})}\BibitemShut {NoStop}%
\bibitem [{\citenamefont {Knowlton}\ and\ \citenamefont
  {Starling}(1912)}]{Knowlton12}%
  \BibitemOpen
  \bibfield  {author} {\bibinfo {author} {\bibfnamefont {F.~P.}\ \bibnamefont
  {Knowlton}}\ and\ \bibinfo {author} {\bibfnamefont {E.~H.}\ \bibnamefont
  {Starling}},\ }\href@noop {} {\bibfield  {journal} {\bibinfo  {journal} {The
  Journal of Physiology}\ }\textbf {\bibinfo {volume} {44}},\ \bibinfo {pages}
  {206} (\bibinfo {year} {1912})}\BibitemShut {NoStop}%
\bibitem [{\citenamefont {Conrad}(1969)}]{Conrad69}%
  \BibitemOpen
  \bibfield  {author} {\bibinfo {author} {\bibfnamefont {W.~A.}\ \bibnamefont
  {Conrad}},\ }\href@noop {} {\bibfield  {journal} {\bibinfo  {journal} {{IEEE
  Transactions on Biomedical Engineering}}\ }\textbf {\bibinfo {volume}
  {{BME-16}}},\ \bibinfo {pages} {284} (\bibinfo {year} {1969})}\BibitemShut
  {NoStop}%
\bibitem [{\citenamefont {Kamm}\ and\ \citenamefont {Shapiro}(1979)}]{Kamm79}%
  \BibitemOpen
  \bibfield  {author} {\bibinfo {author} {\bibfnamefont {R.~D.}\ \bibnamefont
  {Kamm}}\ and\ \bibinfo {author} {\bibfnamefont {A.~H.}\ \bibnamefont
  {Shapiro}},\ }\href@noop {} {\bibfield  {journal} {\bibinfo  {journal}
  {Journal of Fluid Mechanics}\ }\textbf {\bibinfo {volume} {95}},\ \bibinfo
  {pages} {1} (\bibinfo {year} {1979})}\BibitemShut {NoStop}%
\bibitem [{\citenamefont {Jensen}\ and\ \citenamefont {Heil}(2003)}]{Jensen03}%
  \BibitemOpen
  \bibfield  {author} {\bibinfo {author} {\bibfnamefont {O.~E.}\ \bibnamefont
  {Jensen}}\ and\ \bibinfo {author} {\bibfnamefont {M.}~\bibnamefont {Heil}},\
  }\href@noop {} {\bibfield  {journal} {\bibinfo  {journal} {Journal of Fluid
  Mechanics}\ }\textbf {\bibinfo {volume} {481}},\ \bibinfo {pages} {235}
  (\bibinfo {year} {2003})}\BibitemShut {NoStop}%
\bibitem [{\citenamefont {Bertram}(2008)}]{Bertram08}%
  \BibitemOpen
  \bibfield  {author} {\bibinfo {author} {\bibfnamefont {C.~D.}\ \bibnamefont
  {Bertram}},\ }\href@noop {} {\bibfield  {journal} {\bibinfo  {journal}
  {Respiratory Physiology \& Neurobiology}\ }\textbf {\bibinfo {volume}
  {163}},\ \bibinfo {pages} {256} (\bibinfo {year} {2008})}\BibitemShut
  {NoStop}%
\bibitem [{\citenamefont {Stewart}\ \emph {et~al.}(2009)\citenamefont
  {Stewart}, \citenamefont {Waters},\ and\ \citenamefont {Jensen}}]{Stewart09}%
  \BibitemOpen
  \bibfield  {author} {\bibinfo {author} {\bibfnamefont {P.~S.}\ \bibnamefont
  {Stewart}}, \bibinfo {author} {\bibfnamefont {S.~L.}\ \bibnamefont {Waters}},
  \ and\ \bibinfo {author} {\bibfnamefont {O.~E.}\ \bibnamefont {Jensen}},\
  }\href {\doibase 10.1016/j.euromechflu.2009.03.002} {\bibfield  {journal}
  {\bibinfo  {journal} {European Journal of Mechanics-B/Fluids}\ }\textbf
  {\bibinfo {volume} {28}},\ \bibinfo {pages} {541} (\bibinfo {year}
  {2009})}\BibitemShut {NoStop}%
\bibitem [{\citenamefont {Low}\ and\ \citenamefont {Chew}(1991)}]{Low91}%
  \BibitemOpen
  \bibfield  {author} {\bibinfo {author} {\bibfnamefont {H.~T.}\ \bibnamefont
  {Low}}\ and\ \bibinfo {author} {\bibfnamefont {Y.~T.}\ \bibnamefont {Chew}},\
  }\href@noop {} {\bibfield  {journal} {\bibinfo  {journal} {{Medical \&
  Biological Engineering \& Computing}}\ }\textbf {\bibinfo {volume} {29}},\
  \bibinfo {pages} {217} (\bibinfo {year} {1991})}\BibitemShut {NoStop}%
\bibitem [{\citenamefont {Tubaldi}\ \emph {et~al.}(2016)\citenamefont
  {Tubaldi}, \citenamefont {Amabili},\ and\ \citenamefont
  {Pa{\"\i}doussis}}]{Tubaldi2016}%
  \BibitemOpen
  \bibfield  {author} {\bibinfo {author} {\bibfnamefont {E.}~\bibnamefont
  {Tubaldi}}, \bibinfo {author} {\bibfnamefont {M.}~\bibnamefont {Amabili}}, \
  and\ \bibinfo {author} {\bibfnamefont {M.~P.}\ \bibnamefont
  {Pa{\"\i}doussis}},\ }\href@noop {} {\bibfield  {journal} {\bibinfo
  {journal} {Journal of Sound and Vibration}\ }\textbf {\bibinfo {volume}
  {371}},\ \bibinfo {pages} {252} (\bibinfo {year} {2016})}\BibitemShut
  {NoStop}%
\bibitem [{\citenamefont {Tsigklifis}\ and\ \citenamefont
  {Lucey}(2017)}]{tsigklifis2017}%
  \BibitemOpen
  \bibfield  {author} {\bibinfo {author} {\bibfnamefont {K.}~\bibnamefont
  {Tsigklifis}}\ and\ \bibinfo {author} {\bibfnamefont {A.~D.}\ \bibnamefont
  {Lucey}},\ }\href@noop {} {\bibfield  {journal} {\bibinfo  {journal} {Journal
  of Fluid Mechanics}\ }\textbf {\bibinfo {volume} {820}},\ \bibinfo {pages}
  {370} (\bibinfo {year} {2017})}\BibitemShut {NoStop}%
\bibitem [{\citenamefont {Stelios}\ \emph {et~al.}(2019)\citenamefont
  {Stelios}, \citenamefont {Qin}, \citenamefont {Shan},\ and\ \citenamefont
  {Mathioulakis}}]{Stelios19}%
  \BibitemOpen
  \bibfield  {author} {\bibinfo {author} {\bibfnamefont {S.}~\bibnamefont
  {Stelios}}, \bibinfo {author} {\bibfnamefont {S.}~\bibnamefont {Qin}},
  \bibinfo {author} {\bibfnamefont {F.}~\bibnamefont {Shan}}, \ and\ \bibinfo
  {author} {\bibfnamefont {D.}~\bibnamefont {Mathioulakis}},\ }\href@noop {}
  {\bibfield  {journal} {\bibinfo  {journal} {Meccanica}\ }\textbf {\bibinfo
  {volume} {54}},\ \bibinfo {pages} {779} (\bibinfo {year} {2019})}\BibitemShut
  {NoStop}%
\bibitem [{\citenamefont {Amabili}\ \emph
  {et~al.}(2020{\natexlab{a}})\citenamefont {Amabili}, \citenamefont
  {Balasubramanian}, \citenamefont {Ferrari}, \citenamefont {Franchini},
  \citenamefont {Giovanniello},\ and\ \citenamefont {Tubaldi}}]{amabili2020b}%
  \BibitemOpen
  \bibfield  {author} {\bibinfo {author} {\bibfnamefont {M.}~\bibnamefont
  {Amabili}}, \bibinfo {author} {\bibfnamefont {P.}~\bibnamefont
  {Balasubramanian}}, \bibinfo {author} {\bibfnamefont {G.}~\bibnamefont
  {Ferrari}}, \bibinfo {author} {\bibfnamefont {G.}~\bibnamefont {Franchini}},
  \bibinfo {author} {\bibfnamefont {F.}~\bibnamefont {Giovanniello}}, \ and\
  \bibinfo {author} {\bibfnamefont {E.}~\bibnamefont {Tubaldi}},\ }\href@noop
  {} {\bibfield  {journal} {\bibinfo  {journal} {Journal of the Mechanical
  Behavior of Biomedical Materials}\ ,\ \bibinfo {pages} {103804}} (\bibinfo
  {year} {2020}{\natexlab{a}})}\BibitemShut {NoStop}%
\bibitem [{\citenamefont {Amabili}\ \emph
  {et~al.}(2020{\natexlab{b}})\citenamefont {Amabili}, \citenamefont
  {Balsubramanian}, \citenamefont {Bozzo}, \citenamefont {Breslavsky},
  \citenamefont {Ferrari}, \citenamefont {Franchini}, \citenamefont
  {Giovanniello},\ and\ \citenamefont {Pogue}}]{Amabili20}%
  \BibitemOpen
  \bibfield  {author} {\bibinfo {author} {\bibfnamefont {M.}~\bibnamefont
  {Amabili}}, \bibinfo {author} {\bibfnamefont {P.}~\bibnamefont
  {Balsubramanian}}, \bibinfo {author} {\bibfnamefont {I.}~\bibnamefont
  {Bozzo}}, \bibinfo {author} {\bibfnamefont {I.~D.}\ \bibnamefont
  {Breslavsky}}, \bibinfo {author} {\bibfnamefont {G.}~\bibnamefont {Ferrari}},
  \bibinfo {author} {\bibfnamefont {G.}~\bibnamefont {Franchini}}, \bibinfo
  {author} {\bibfnamefont {F.}~\bibnamefont {Giovanniello}}, \ and\ \bibinfo
  {author} {\bibfnamefont {C.}~\bibnamefont {Pogue}},\ }\href@noop {}
  {\bibfield  {journal} {\bibinfo  {journal} {{Physical Review X}}\ }\textbf
  {\bibinfo {volume} {10}},\ \bibinfo {pages} {011015} (\bibinfo {year}
  {2020}{\natexlab{b}})}\BibitemShut {NoStop}%
\bibitem [{\citenamefont {Heil}\ and\ \citenamefont {Hazel}(2006)}]{heil2006}%
  \BibitemOpen
  \bibfield  {author} {\bibinfo {author} {\bibfnamefont {M.}~\bibnamefont
  {Heil}}\ and\ \bibinfo {author} {\bibfnamefont {A.~L.}\ \bibnamefont
  {Hazel}},\ }in\ \href@noop {} {\emph {\bibinfo {booktitle} {Fluid-structure
  interaction}}},\ \bibinfo {editor} {edited by\ \bibinfo {editor}
  {\bibfnamefont {M.}~\bibnamefont {Sch\"afer}}\ and\ \bibinfo {editor}
  {\bibfnamefont {H.-J.}\ \bibnamefont {Bungartz}}}\ (\bibinfo  {publisher}
  {Springer},\ \bibinfo {year} {2006})\ pp.\ \bibinfo {pages}
  {19--49}\BibitemShut {NoStop}%
\bibitem [{\citenamefont {Whittaker}\ \emph {et~al.}(2010)\citenamefont
  {Whittaker}, \citenamefont {Heil}, \citenamefont {Jensen},\ and\
  \citenamefont {Waters}}]{whittaker2010}%
  \BibitemOpen
  \bibfield  {author} {\bibinfo {author} {\bibfnamefont {R.~J.}\ \bibnamefont
  {Whittaker}}, \bibinfo {author} {\bibfnamefont {M.}~\bibnamefont {Heil}},
  \bibinfo {author} {\bibfnamefont {O.~E.}\ \bibnamefont {Jensen}}, \ and\
  \bibinfo {author} {\bibfnamefont {S.~L.}\ \bibnamefont {Waters}},\
  }\href@noop {} {\bibfield  {journal} {\bibinfo  {journal} {Quarterly Journal
  of Mechanics and Applied Mathematics}\ }\textbf {\bibinfo {volume} {63}},\
  \bibinfo {pages} {465} (\bibinfo {year} {2010})}\BibitemShut {NoStop}%
\bibitem [{\citenamefont {Walters}\ \emph {et~al.}(2017)\citenamefont
  {Walters}, \citenamefont {Heil},\ and\ \citenamefont
  {Whittaker}}]{Walters2017}%
  \BibitemOpen
  \bibfield  {author} {\bibinfo {author} {\bibfnamefont {M.~C.}\ \bibnamefont
  {Walters}}, \bibinfo {author} {\bibfnamefont {M.}~\bibnamefont {Heil}}, \
  and\ \bibinfo {author} {\bibfnamefont {R.~J.}\ \bibnamefont {Whittaker}},\
  }\href@noop {} {\bibfield  {journal} {\bibinfo  {journal} {{Quarterly Journal
  of Mechanics and Applied Mathematics}}\ }\textbf {\bibinfo {volume} {71}},\
  \bibinfo {pages} {47} (\bibinfo {year} {2017})}\BibitemShut {NoStop}%
\bibitem [{\citenamefont {Mandre}\ and\ \citenamefont
  {Mahadevan}(2010)}]{Mandre2010}%
  \BibitemOpen
  \bibfield  {author} {\bibinfo {author} {\bibfnamefont {S.}~\bibnamefont
  {Mandre}}\ and\ \bibinfo {author} {\bibfnamefont {L.}~\bibnamefont
  {Mahadevan}},\ }\href@noop {} {\bibfield  {journal} {\bibinfo  {journal}
  {{Proceedings of the Royal Society A}}\ }\textbf {\bibinfo {volume} {466}},\
  \bibinfo {pages} {141} (\bibinfo {year} {2010})}\BibitemShut {NoStop}%
\end{thebibliography}%
%

\end{document}

\title{\vspace{-2cm} {\Large Supplementary Materials for}\\[0mm] 
	\textbf{Resonances in pulsatile channel flow with an elastic wall}
}

\author[1,2]{Duo Xu}
\author[3]{Matthias Heil}
\author[2]{Thomas Seeb{\"o}ck}
\author[1,2]{Marc Avila}

\affil[1]{{\small University of Bremen, Center of Applied Space Technology and Microgravity (ZARM), 28359 Bremen, Germany}}
\affil[2]{{\small Friedrich-Alexander-Universit{\"a}t Erlangen-N{\"u}rnberg, 91058 Erlangen, Germany}}
\affil[3]{{\small University of Manchester, School of Mathematics, Manchester, M13 9PL United Kingdom}}
\date{\today}

\maketitle

\section*{Model Derivation}

The assumption that the flow in the upstream and downstream rigid
segments is {dominated by its horizontal component} implies that  the
pressure gradient in these segments is {approximately} spatially constant, ${\partial
  p}/{\partial x}=G(t)$. Using that the fluid pressure under the
membrane is $p_{\rm m}=kY(t)$, the momentum equation ${\partial
  u}/{\partial t} = -{\partial p}/{\partial x} + {\partial^2
  u}/{\partial y^2}$ (equation~(5) in the paper) reads
\be
\label{mom}
\frac{\partial u_{\rm u/d}(y,t)}{\partial t} =
-G_{\rm u/d}(t) +
\frac{\partial^2 u_{\rm u/d}(y,t)}{\partial y^2}
\ee
with 
\be
\label{eq:G_u}
G_{\rm u}(t)=\frac{k Y(t) - p_{\rm in}(t)}{l_{\rm u}}
\ee
and
\be
\label{eq:G_d}
G_{\rm d}(t)=-\frac{k Y(t)}{l_{\rm d}}
\ee
for the upstream and downstream segments, respectively. The flows in the two rigid segments are coupled by
mass conservation which requires that the
rate of change in the volume of the elastic segment 
is matched by the net volume flux from the rigid segments
\be
\int_0^{l_{\rm m}} \frac{\partial y_{\rm m}}{\partial t} \ {\rm d}x =
 \int_0^1  (u_{\rm u} - u_{\rm d} ) \ {\rm d}y.
\ee
For the membrane shape assumed in the paper (equation~(6))
\be
y_{\rm m}(x,t) = 1+ Y(t)\, M(x), 
\ee
with $M(x) = 4 (x/l_{\rm m})(1-(x/l_{\rm  m}))$, we obtain 
\be
\int_0^{l_{\rm m}} \frac{\partial y_{\rm m}}{\partial t} \ {\rm d}x =
\frac{{\rm d}Y}{{\rm d}t} \int_0^{l_{\rm m}} M(x) \ {\rm d}x  = 
\frac{2}{3} \ l_{\rm m} \ \frac{{\rm d}Y}{{\rm d}t},
\ee
so that
\be
\label{conti}
\frac{2}{3} \ l_{\rm m}  \ \frac{{\rm d}Y}{{\rm d}t}
= \int_0^1  (u_{\rm u} - u_{\rm d} ) \ {\rm d}y.
\ee

\subsection*{Analytical solution}

To explore under what conditions the system displays resonant
behavior we consider the case of a purely oscillatory forcing, $p_{\rm in}(t) =
\widehat{P} \exp({\rm i}\alpha^2 t)$ and assume a time-periodic response such
that $Y(t) = \widehat{Y} \exp({\rm i}\alpha^2 t)$. Inserting this ansatz into \eqref{mom} 
and solving the resulting ordinary differential equation, subject to
the no-slip boundary conditions, we obtain the solution
\be
\label{womm_sol}
  u(y,t)_{\rm u/d}=\left(1-\cosh(\sqrt{{\rm i}}\alpha y)  + \dfrac{\cosh(\sqrt{{\rm i}}\alpha)-1}{\sinh(\sqrt{{\rm i}}\alpha)} \sinh(\sqrt{{\rm i}}\alpha y)\right)\, \dfrac{{\rm i}}{\alpha^2}\widehat{G}_{\rm u/d} \exp({\rm i}\alpha^2 t),
\ee
where $\widehat{G}_{\rm u}  =\widehat{Y}k/l_{\rm u}-\widehat{P}/l_{\rm u}$ and $\widehat{G}_{\rm d} =-\widehat{Y}k/l_{\rm d}$ and, for the upstream and downstream segments, respectively. Substituting these two solutions into the right-hand-side of equation~\eqref{conti} and integrating we obtain
\[
\int_0^1  (u_{\rm u} - u_{\rm d} ) \ {\rm d}y = \frac{{\rm i}}{ \alpha^2}\left(1 + \Gamma(\alpha)\right)\left(k \frac{l_{\rm u} + l_{\rm d}}{l_{\rm u} l_{\rm d}}\widehat{Y}-\frac{\widehat{P}}{l_{\rm u}} \right)\exp({\rm i}\alpha^2 t),
\]
where $\Gamma(\alpha) = \dfrac{2}{\sqrt{\rm i} \alpha}\dfrac{1-\cosh(\sqrt{{\rm i}}\alpha)}{\sinh(\sqrt{{\rm i}}\alpha)}$. By inserting $Y(t) = \widehat{Y} \exp({\rm i}\alpha^2 t)$ into the left-hand-side and differentiating we obtain
\be
\label{amp_eq}
\frac{2}{3} \ \alpha^4 \ l_{\rm m} \ \widehat{Y} =  \left(1 + \Gamma(\alpha)\right)\left(k \frac{l_{\rm u} + l_{\rm d}}{l_{\rm u} l_{\rm d}}\widehat{Y}-\frac{\widehat{P}}{l_{\rm u}} \right).
\ee
The solution reads
\be
\label{comp_amp_sol}
\widehat{Y} = \frac{-(1+\Gamma(\alpha))\widehat{P}}
{l_{\rm u}\left(\dfrac{2}{3} \alpha^4 l_{\rm m}- \left(1 + \Gamma(\alpha)\right) k \dfrac{l_{\rm u} + l_{\rm d}}{l_{\rm u} l_{\rm d}}   \right)} = \frac{3}
{2 l_{\rm u}l_{\rm m}}\frac{-(1+\Gamma(\alpha))\widehat{P}}{\alpha^4 - \left(1 + \Gamma(\alpha)\right)\alpha_{\rm eig}^4}. 
\ee
and the amplitude of the oscillations is
\be
\label{amp_sol}
|\widehat{Y}|= 
 \frac{18 l \Rey A}{l_{\rm u}l_{\rm m}}\sqrt{\frac{\left(1+\Re[\Gamma(\alpha)]\right)^2 +\Im[\Gamma(\alpha)]^2}{\left(\alpha^4 - \left(1+\Re[\Gamma(\alpha)]\right)\alpha_{\rm eig}^4\right)^2 + \left(\alpha_{\rm eig}^4\Im[\Gamma(\alpha)]\right)^2}},
\ee
where $\Re[\Gamma(\alpha)]$ and $\Im[\Gamma(\alpha)]$ denote the real
and the imaginary parts of $\Gamma(\alpha)$, respectively, and $\widehat{P} =
-12 \ l \ A \ \Rey \ {\rm i}$ (cf.\ equation (2) in the paper). The
frequency $\alpha^2_{\rm max}$ at which the system has its maximum
response is found by maximizing equation~\eqref{amp_sol} with respect to
$\alpha^2$, which must be done numerically.

Differentiating equation~\eqref{conti} with respect to time and inserting 
$\partial u_{\rm u/d}/\partial t$
from \eqref{mom} yields equation (9)
in the paper
\be
\label{eq9paper}
\frac{2}{3}l_{\rm m} \frac{{\rm d}^2 Y}{{\rm d}t^2} +
k \frac{l_{\rm u} + l_{\rm d}}{l_{\rm u} l_{\rm d}} \ Y
+ \left.
\left( \frac{\partial u_{\rm d}}{\partial y} -
       \frac{\partial u_{\rm u}}{\partial y} \right)
\right|_{y=0}^1 = \frac{p_{\rm in}(t)}{l_{\rm u}},
\ee
where
%\[
%\left. \frac{\partial u_{\rm d}}{\partial y}\right|_{y=0}^1=2\left.\frac{\partial u_{\rm d}}{\partial y}\right|_{y=0} = \dfrac{2{\rm i}\sqrt{{\rm i}}}{\alpha}\dfrac{\cosh(\sqrt{{\rm i}}\alpha)-1}{\sinh(\sqrt{{\rm i}}\alpha)}\, \left(\frac{-\widehat{Y} k}{l_{\rm d}}\right) \exp({\rm i}\alpha^2 t)
%\]
%and
%\[
%\left. \frac{\partial u_{\rm u}}{\partial y}\right|_{y=0}^1=2\left.\frac{\partial u_{\rm u}}{\partial y}\right|_{y=0} =  \dfrac{2{\rm i}\sqrt{{\rm i}}}{\alpha}\dfrac{\cosh(\sqrt{{\rm i}}\alpha)-1}{\sinh(\sqrt{{\rm i}}\alpha)}\,\left(\frac{\widehat{Y} k}{l_{\rm u}}-\frac{\widehat{P}}{l_{\rm u}}\right) \exp({\rm i}\alpha^2 t).
%\]
%and so 
\[
\left. \left(\frac{\partial u_{\rm d}}{\partial y} - \frac{\partial u_{\rm u}}{\partial y}\right) \right|_{y=0}^1 = \Gamma(\alpha) \left(\frac{\widehat{P}}{l_{\rm u}} -k \frac{l_{\rm u} + l_{\rm d}}{l_{\rm u} l_{\rm d}}\widehat{Y} \right)\exp({\rm i}\alpha^2 t).
\]
This shows that the factor $\Gamma(\alpha)$ arises solely due to the viscous shear stresses {on the channel walls.} In the absence of viscous shear stresses the collapsible channel responds as an undamped forced harmonic oscillator (see also equation~\eqref{amp_sol}). The viscous effects can be decomposed into three distinct (frequency-depending) contributions:
\begin{enumerate}
\item The $\Gamma(\alpha)\widehat{P}$-term modifies the effective amplitude and phase of the forcing;
\item The real part of the $\Gamma(\alpha)\widehat{Y}$-term contributes to the spring stiffness;
\item The imaginary part of the $\Gamma(\alpha)\widehat{Y}$-term provides the viscous damping.
\end{enumerate}
We thus refer to the effect of the viscous shear stress term as non-conventional damping. 
%The real part of $\Gamma_\alpha$ is negative, it tends to $-1$ as $\alpha\rightarrow 0$, and to $0$ (proportionally to $\alpha^{-1}$) as $\alpha\rightarrow \infty$. The imaginary part of $\Gamma_\alpha$ is positive and tends to $0$ in both limits (proportionally to $\alpha$ and $\alpha^{-1}$, respectively). It achives a maximum of $0.4172$ at $\alpha=3.19$.

\subsection*{Physical interpretation}

{
Noting that $p_{\mathrm m}=kY(t)$ is the pressure in the central section, with a little
re-arrangement, equation (9) in the paper can be rewritten as
\begin{equation}
\frac{2}{3}l_{\mathrm m} \frac{\mathrm{d}^2 Y}{\mathrm{d} t^2} =
\left[\frac{1}{l_{\mathrm u}} (p_\mathrm{in} - kY) -
  \left.\frac{\mathrm{d}u_{\mathrm d}}{\mathrm{d}y} \right\vert_0^1 \right] -
\left[\frac{1}{l_{\mathrm d}} (kY-0) -
  \left.\frac{\mathrm{d}u_{\mathrm u}}{\mathrm{d}y}\right|_0^1\right].
\end{equation}
This shows that the difference in the pressure gradients in the upstream and downstream segments
drives an acceleration in the net flow away from the centre, which
must be balanced by the change in volume of the elastic section.
}

\subsection*{Additional physical mechanisms}

{
In our numerical computations and analytical model, we assume that the membrane is massless. 
The incorporation of membrane mass would have two additional effects.
\begin{description}
  \item[(i)] It would provide ``added mass'' to the inertia that 
is currently provided exclusively by the fluid; this would change
(lower) the natural frequencies of the system, see e.g.~[21].
\item[(ii)] It would
allow the membrane to perform oscillations with its ``own'' natural
frequency, even without the presence of the fluid. These
purely solid-based oscillations would be modified by the presence
of the fluid (which would now provide ``added mass'' to the solid) and
add further resonances and the potential for flutter-like
instabilities to the system, see e.g.~[22].
\end{description}
}

\pagebreak

%%%% parameter table %%%%%%
\definecolor{aqua}{rgb}{0.0, 1.0, 1.0}
{\renewcommand{\arraystretch}{1.25}
	\begin{table}[htb]
		\centering
		\begin{tabular}{l l l l|l l l| c}
			\toprule
			$\sigma_0$ & $T$ & $Re$ & $h$ & $l_\textrm{u}$ & $l_\textrm{m}$ & $l_\textrm{d}$ & symbol \\ 
			\hline
			$10^3$ & $2.5\times10^6-2\times10^9$ & $0.25-200$ & $0.01$ & $5$ & $10$ & $10$ &  {\LARGE\textcolor{Orange}{$\bullet$}} \\ 
			$10^3$ & $10^7-2\times10^{10}$ & $1-200$ & $0.01$ & $5$ & $10$ & $10$ &  \textcolor{aqua}{$\filledmedtriangleleft$}  \\ 
			$10^3$ & $5\times10^5-10^9$ & $0.10-200$ & $0.01$ & $5$ & $10$ & $10$ &  \textcolor{blue}{$\filledmedtriangleup$}  \\ 	
			$10^3$ & $5\times10^7-10^{10}$ & $1-200$ & $0.01$ & $5$ & $10$ & $10$ & \textcolor{green}{$\filledmedtriangledown$} \\ %\hline
			
			$10^3$ & $10^{10}$ & $100$ & $0.01$ & $5$ & $10$ & $10$ & \textcolor{Black}{$\filledmedtriangleright$}  \\ 
			$10^4$ & $10^{10}$ & $100$ & $0.01$ & $5$ & $10$ & $10$ & \textcolor{Black}{$\filledmedtriangleleft$}   \\ 				
			$10^3$ & $10^{11}$ & $100$ & $0.01$ & $5$ & $10$ & $10$ &  \textcolor{Black}{$\filledmedtriangleright$}   \\ 				
			$10^4$ & $10^{11}$ & $100$ & $0.01$ & $5$ & $10$ & $10$ &  \textcolor{Black}{$\filledmedtriangleleft$}  \\ %\hline			
			\hline
			$10^3$ & $10^{9}$ & $100$ & $0.005$ & $5$ & $10$ & $10$ &  {\LARGE\textcolor{Black}{$\bullet$}}   \\ 				
			$10^3$ & $10^{9}$ & $100$ & $0.02$ & $5$ & $10$ & $10$ &  {\LARGE\textcolor{Black}{$\circ$}}   \\ %\hline						
			$10^3$ & $10^{9}$ & $100$ & $0.01$ & $5$ & $5$ & $10$ & \textcolor{Red}{$\filledmedsquare$} \\ 
			$10^3$ & $10^{9}$ & $100$ & $0.01$ & $5$ & $8$ & $10$ & \textcolor{Green}{$\filledmedtriangleup$} \\ 
			$10^3$ & $10^{9}$ & $100$ & $0.01$ & $5$ & $10$ & $10$ &  {\LARGE\textcolor{aqua}{$\bullet$}} \\ 
			$10^3$ & $10^{9}$ & $100$ & $0.01$ & $10$ & $10$ & $10$ &  \textcolor{blue}{$\filledmedtriangledown$} \\ 
			$10^3$ & $10^{9}$ & $100$ & $0.01$ & $5$ & $10$ & $15$ &  \textcolor{VioletRed}{$\filledmedtriangleright$} \\ 		
			$10^3$ & $10^{9}$ & $100$ & $0.01$ & $10$ & $10$ & $15$ &  {\textcolor{Orange}{$\filledmedtriangleleft$}} \\ 
			\bottomrule
		\end{tabular}
		
		\caption{\label{tab:parameter} 
		Simulation parameters for the data points displayed in figure~\ref{fig:geometry}. In all cases the amplitude is $A=1$, the external pressure is {$p_{\rm ext}=10^{-4}$ (except for the cases  with $\sigma=10^4$, where $p_{\rm ext}=10^{-5}$)} and the dimensionless frequency is in the range $0.05\leqslant\alpha^2\leqslant 2250$. The first eight rows show the parameters corresponding to the data points displayed in figure~5(b) of the paper (in which the dimensions of the channel are kept fixed). }
	\end{table}

\newpage

\begin{figure}
	\centering \includegraphics[width=0.6\linewidth, trim={0cm 0cm 0cm 0cm}, clip]{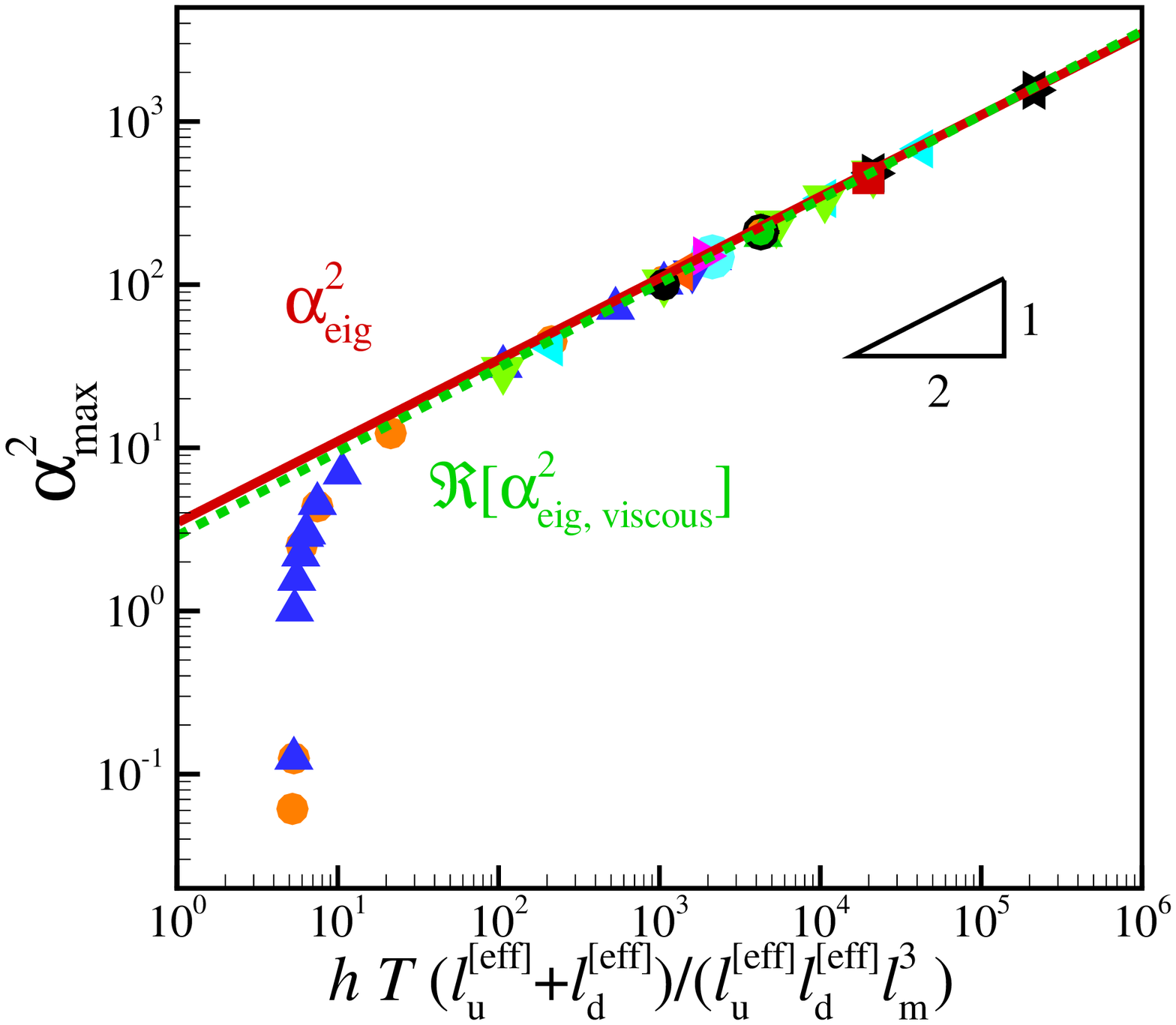}
	\caption{\label{fig:geometry} Frequency $\alpha^2_{\text{max}}$  at which the system has its maximum the maximum as a function of $h\,T (l^{\text{[eff]}}_{\rm u}+l^{\text{[eff]}}_{\rm d})/(l^{\text{[eff]}}_{\rm u}l^{\text{[eff]}}_{\rm d} l_{\rm m}^3)$. The symbols show the results of our numerical simulations (the parameter values are listed in {Table}~\ref{tab:parameter}). The red solid line denotes $\alpha^2_{\rm eig}$ from the model prediction, and the green dotted line corresponds to the real part of the (complex) viscous eigenfrequency $\alpha^2_{\rm eig,viscous}$, which is obtained by numerically solving the viscous eigenvalue problem given by equation~(9) with $p_{\rm in}=0$.}
\end{figure}

\clearpage
\newpage

\begin{figure}
	\centering \includegraphics[width=0.85\linewidth, trim={0cm 0cm 0cm 0cm}, clip]{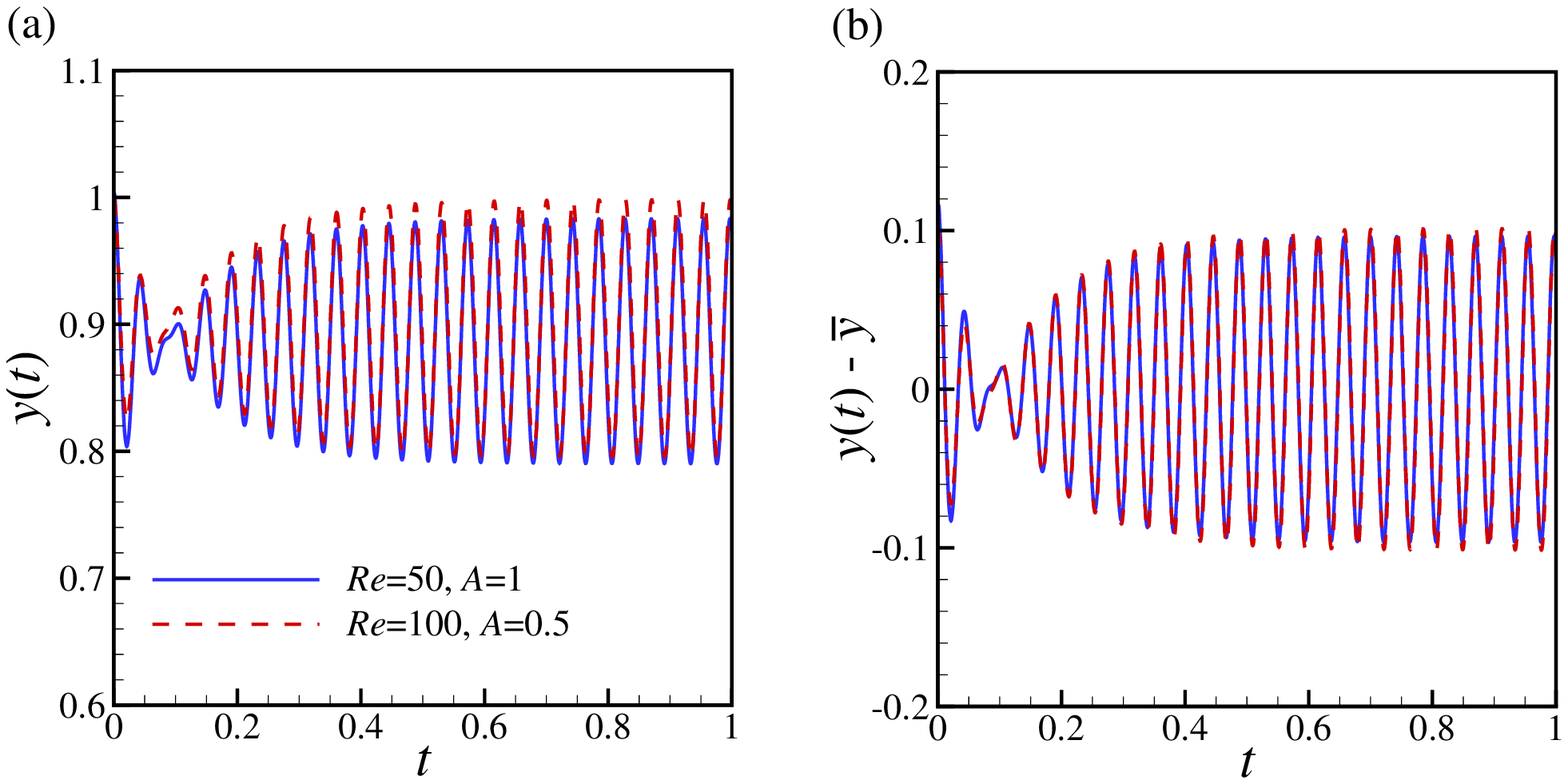}\\
	\centering \includegraphics[width=0.85\linewidth, trim={0cm 0cm 0cm 0cm}, clip]{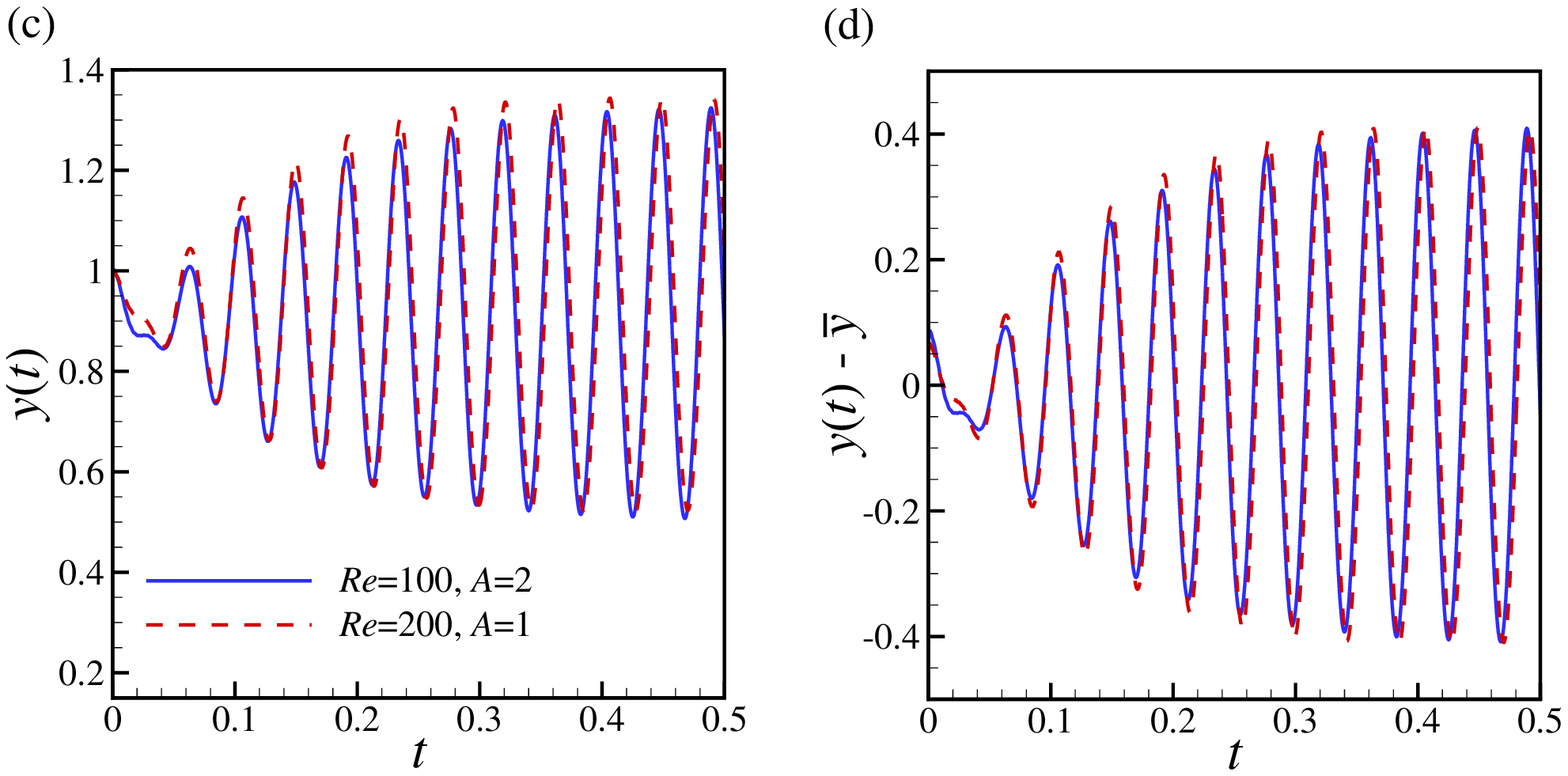}
	\caption{\label{fig:pulsation_amplitude}  Influence of amplitude of the pressure pulsation ($A$) on the membrane oscillation at $\alpha^2=148$, $T=10^9$, {$p_\text{ext}=10^{-5}$,} $\sigma_0=10^3$, $l_\textrm{u}=5$, $l_\textrm{m}=10$, $l_\textrm{d}=10$  and $h=0.01$. (a) Time trace of the membrane mid-point $y(t)$ for $Re=100, A=0.5$ and $Re=50, A=1$. (b) The same data of (a) after subtracting the time-averaged displacement. (c) Time trace of the membrane mid-point $y(t)$ for $Re=100, A=2$ and $Re=200, A=1$. (d) The same data of (c) after subtracting the time-averaged displacement. } 
\end{figure}

\bibliographystyle{plain}
\nobibliography{FSI}